\documentclass[12pt, a4paper, parskip, DIV=15]{scrartcl}
\usepackage[english]{babel}
\usepackage[utf8]{inputenc}
\usepackage[T1]{fontenc}  
\usepackage{lmodern, textcomp, booktabs}
\usepackage[hidelinks]{hyperref}
\usepackage{bm}
\usepackage[round]{natbib}
\usepackage{authblk, doi}
\usepackage{grffile}  
\usepackage[separate-uncertainty=true]{siunitx}
\usepackage{amsmath, xspace}
\usepackage[format=plain, indention=0.5cm, font=small]{caption}
\usepackage{subcaption}
\usepackage{cleveref}
\usepackage{csquotes}
\usepackage{graphicx}
\graphicspath{{figures/}}

\catcode`\^^M=10

\title{\LARGE Numerical simulations of seismo-acoustic nuisance patterns from an induced M1.8 earthquake in the Helsinki, southern Finland, metropolitan area}
\author[1]{Lukas Krenz}
\author[1]{Sebastian Wolf}
\author[2]{Gregor Hillers}
\author[3,4.*]{Alice-Agnes Gabriel}
\author[1]{Michael Bader}

\affil[1]{TUM School of Computation, Information and Technology; Department of Computer Science; Technical University of Munich; Munich, Germany}
\affil[2]{Institute of Seismology, University of Helsinki, Helsinki, Finland}
\affil[3]{Scripps Institution of Oceanography, UC San Diego, La Jolla, CA, USA}
\affil[4]{Department of Earth and Environmental Sciences, Ludwig-Maximilians-Universität München, Munich, Germany}
\affil[*]{contact: algabriel@ucsd.edu}
\date{{\vspace{-0.5cm}\small \today}}

\hypersetup{pdfauthor={L. Krenz et al.}, pdftitle={Modelled seismo-acoustic nuisance patterns in Helsinki}}

\newcommand{\ind}[1]{_\text{#1}}

\newcommand{\ML}{\ensuremath{M\ind L}}
\newcommand{\Mw}{\ensuremath{M\ind w}}

\begin{document}

\maketitle
\begin{abstract}\noindent
Seismic waves can couple with the atmosphere and generate sound waves.
The influence of faulting mechanisms on earthquake sound patterns provides opportunities for earthquake source characterization.
Sound radiated from earthquakes can be perceived as disturbing, even at low ground shaking levels, which can negatively impact the social acceptance of geoengineering applications.
Motivated by consistent reports of felt and heard disturbances associated with the week-long stimulation of a \SI{6}{\km}-deep geothermal system in 2018 below the Otaniemi district of Espoo, Helsinki, we conduct fully coupled 3D numerical simulations of wave propagation in the solid Earth and the atmosphere. 
We assess the sensitivity of the ground shaking and audible noise distributions to the source geometry of the induced earthquakes based on the properties of the largest local magnitude \ML1.8 event.
Utilizing recent computational advances and the open-source software SeisSol, we model seismo-acoustic frequencies up to \SI{25}{\Hz}, thereby reaching the lower limit of the audible sound frequency range.
We present synthetic distributions of shaking and audible sounds at the \SIrange[range-phrase={ to }]{50}{100}{\m} scale across a $\SI{12}{\km} \times \SI{12}{\km}$ area
and discuss implications for better understanding seismic nuisances in metropolitan regions. 
In five 3D coupled elastic-acoustic scenario simulations that include data on topography and subsurface structure, we analyze the ground velocity and pressure levels of earthquake-generated seismic and acoustic waves.
We show that $S$~waves generate the strongest sound disturbance with sound pressure levels of up to 0.04~Pa.
We use statistical analysis to compare our noise distributions with commonly used empirical relationships.
We find that our 3D synthetic amplitudes are generally smaller than the empirical predictions and that the interaction of the source mechanism-specific radiation pattern and topography can lead to significant non-linear effects.
Our study highlights the complexity and information content of spatially variable audible effects associated with small induced earthquakes on local scales.
\end{abstract}

\section{Introduction}
\label{sec:intro}
\begin{figure}
    \centering
    \includegraphics[scale=0.5]{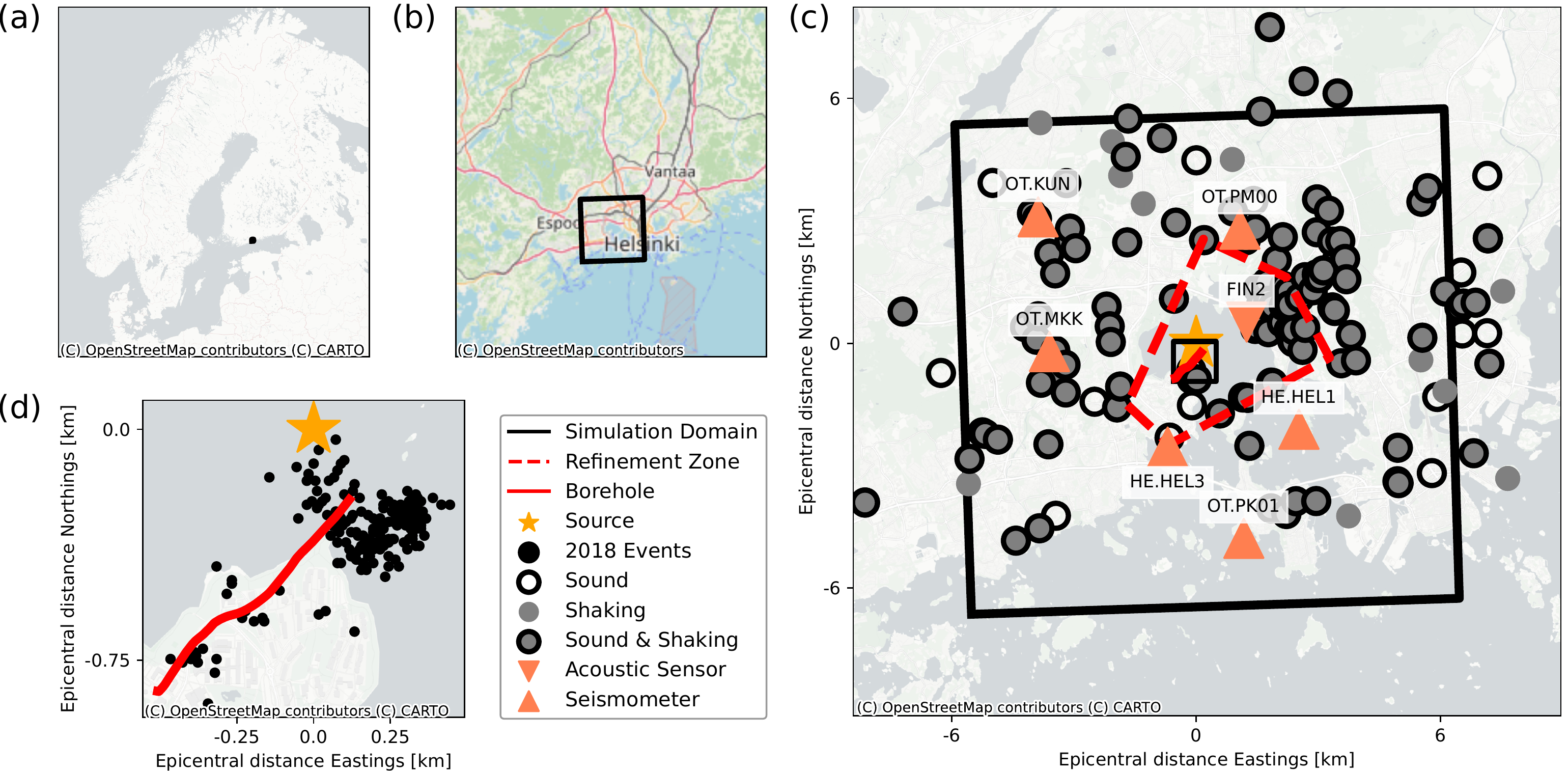}
    \caption{
    (a) Location of the study area in Northern Europe; the black symbol marks the location of the black squares in (b) and (c).
    (b) Location of the study area in the Helsinki metropolitan area. The black square indicates the computational domain, as shown in (c).
    (c) Circles show the locations of all 220 macroseismic reports submitted during the 2018 stimulation period. Their locations are accurate to the street addresses.
    Circles with black outlines indicate reported sound observations. Gray circles (no outline) indicate the reported shaking sensations.
    Gray circles with black outlines represent observations of both, sound and shaking.
    The large black square indicates the $\SI{12}{\km} \times \SI{12}{\km}$ simulation domain. The small black square in the center shows the epicentral area enlarged in (d). The dashed red polygon encloses the source region and neighborhoods to the north-east, from where many macroseismic reports were collected. In this region, we compute the seismo-acoustic wavefields with the highest accuracy.
    The inverted triangle indicates the location of the FIN2 microphone array \citep{Lamb2021}, and the other triangles indicate the selection of seismic stations that we use for data comparison.
    The star indicates the location of the largest induced \ML1.8 event 13 in \citet{hillers_geothermal_2020}. 
    All figures use this location as the origin.
    (d) The 2018 borehole trajectory is indicated in red, as in (c). 
    The borehole is vertical down to a depth of 5000~m and then inclines in the north-east direction.
    Black symbols show 203 manually revised event locations \citep{hillers_geothermal_2020}, which are a small subset of the several thousands of induced earthquakes. 
    All maps use the Webmercator coordinate system, which is different from the map projection used for the simulations.
    }
    \label{fig:overview}
\end{figure}


Induced seismicity is essential to increase the reservoir flow rate in an enhanced geothermal system (EGS).
Hence, the mitigation of damaging earthquake ground motions is important for this and other geoengineering applications.
This is why increasing attention has been paid to the assessment of acceptable ground shaking limits associated with anthropogenic seismicity \citep[e.g.,][]{Brooks2018,Keil2022}.
In contrast, the excitation of potentially disturbing audible effects appears to be of minor \citep{Megies2014} or no concern \citep{majer2012protocol}.
However, audible earthquake signals (rumbling, thuds, roaring, sonic booms, clattering, or muffled sounds) have been reported in epicentral areas of large and small events \citep{Davison1938,Ebel1982,Sylvander2007,hill2011}. 
Earthquake sound patterns have been linked to faulting mechanisms, which implies the possibility of non-instrumental earthquake source characterization from sufficiently dense and complete macroseismic surveys \citep[e.g.,][]{Tosi2000,Mantyniemi2004}.
\par
%
%
The 2018 and 2020 deep geothermal stimulation experiments in the Helsinki metropolitan area have led in total to more than 300 macroseismic reports of felt and heard earthquake effects \citep{Ader2019,hillers_geothermal_2020,Rintamaki2021,Lamb2021}.
This demonstrates that persistent earthquake noise has the potential to negatively affect the public attitude towards stimulation activities \citep{Stauffacher2015}. 
Research into local audible sound excitation mechanisms of small earthquakes can therefore support a smoother implementation of geothermal systems
that are ideally developed near a large consumer base which may express concerns about nuisances associated with vibrations and noise related to the stimulation.
%
\par
Monitoring procedures and intervention protocols, including traffic light systems (TLSs), typically focus on the limitation of earthquake magnitude or other decision variables to prevent shaking levels that are hazardous to infrastructure or pose a nuisance to affected communities \citep{Bommer2006}.
\citet{Baisch2019} recalls implicit assumptions for TLS operations, most notably the belief that real-time responses to stimulated seismicity can prevent an increase in earthquake size after the action.
Examples of TLS failures, such as the post-injection earthquake of the 2006 Basel, Switzerland stimulation \citep{Haring2008}, highlight the dependence of a successful TLS implementation on the employed forecast model.
Forecast models can be based on simple frequency-magnitude relationships but can also be composed of complex systems that include multi-physics feedback mechanisms between the injected fluids and monitored seismicity \citep{Gaucher2015}. The latter can be a key module in an operational strategy that constantly adapts to evolving situations \citep{Grigoli2017,Mignan2017}.
The definition of TLS threshold values, such as tolerable earthquake magnitude, can appear ad hoc \citep{Baisch2019}. Structurally safe shaking levels  may still reach an intensity that is unacceptable to a critical part of the public \citep{Rutqvist2014perception}.
This is addressed by approaches that highlight the need for spatially varying limit definitions instead of single-valued thresholds for a larger area,
and consider probabilistic models of perceptible ground shaking for TLS threshold determination \citep{Schultz2021a,Schultz2021b}.
State of the art mitigation strategies do not consider acoustic emissions or noise generation \citep{Verdon2021}.
This reflects the preconception that ``seismicity usually does not radiate sufficient noise to be audible'' \citep[][]{majer2012protocol},
although this topic is recognized in geothermal energy social acceptance studies \citep[``many residents  \ldots are afraid of noise'';][]{Stauffacher2015}.
\par
Public perception of induced seismic effects, including sound, can be potentially critical to geothermal project management.
This requires a broad knowledge base of the physical mechanisms for sound generation and quantification of these effects to inform tailored mitigation strategies and region-specific solutions.
%
Key questions related to sound generation associated with subsurface seismic sources include whether 
sound excitation is confined to the epicentral area,
whether sound waves are locally generated through seismo-acoustic coupling, or whether secondary sources associated with topography play a dominant role \citep{Arrowsmith2010,Lamb2021},
and how meteorological conditions affect propagation.
These questions are relevant to seismo-acoustic waves at all frequencies, regardless of whether sound waves are audible or not.
\par
Epicentral, local, or diffracted earthquake infrasound, or inaudible sound, with a frequency content below \SI{20}{\Hz} can be excited by the coupling of $P$~waves, $S$~waves, and surface waves
\citep{Mutschlecner2005,Evers2014,hernandezEstimatingGroundMotion2018,shani-kadmiel2010HaitiEarthquake2021}.
Infrasound observations are supported by a mature network of global permanent infrasound sensors \citep{hedlin2012infrasound,Hupe2022}, and temporary network densification \citep{VernonEGU2012} allows locally higher resolution of transient local phenomena \citep{Edwards2014}.
Modeling of seismo-acoustic coupling may include atmospheric propagation effects.
Studies of infrasound propagation from sources to receivers across regional and global distances generally account for the 3D atmospheric structure between the ground and heights up to an altitude of \SI{120}{\km} in the thermosphere.
Ray tracers \citep{Arrowsmith2010}, propagation models based on the parabolic equations of infrasound \citep{waxler2022two}, or other approaches \citep{WaxlerBook2019} include the effects of wind and temperature that also govern dissipation.
\par
Audible earthquake noise patterns excited by small to moderate earthquakes \citep{Sylvander2005,Thouvenot2009,Mantyniemi2022}
on local scales, which are relevant here, are often approximated using relationships between vertical ground motion and induced sound pressure that disregard atmospheric propagation effects \citep{hill1976earthquake,Tosi2000,Lamb2021}.
Compared to infrasound data, audible noise observations are sparse; however, increasingly dense high-frequency sound observations increase the demand for excitation and propagation models with comparable resolution.
%
\par
Macroseismic report patterns across Helsinki \citep{hillers_geothermal_2020} show strong spatial variations at the district or neighborhood scale (\cref{fig:overview}). This highlights the need to accurately model small-scale complexities of seismo-acoustic energy coupling and propagation at frequencies that include at least the lower limit of the human audible frequency range around 20~Hz. 
The Helsinki stimulation was conducted in the southern Finland Fennoscandian Shield environment.
The absence of sedimentary layers means that the \SI{6}{\km} deep stimulated bedrock units crop out and high-frequency seismic energy reaches the surface.
This geological situation leads to recurrently observed felt and heard experiences reported by an urban population unfamiliar with earthquake phenomena.
The models presented here can quantitatively support research into geological, infrastructural, socioeconomic, or other factors that govern public response pattern.
However, the inherent small length scales and multi-physics characteristics pose serious challenges to numerical approaches.
%
\par
Advances in high-performance computing facilitate the investigation of the generation and propagation of a variety of seismo-acoustic signals, including the generation of ultra-low-frequency acoustic $T$ waves in 2D and 3D simulations \citep[e.g.][]{averbuchLongrangeAtmosphericInfrasound2020,cheIlluminatingNorthKorean2022}. However, 3D coupled seismo-acoustic simulations remain computationally challenging. For example, \citet{Lecoulant2019} modeled oceanic $T$ waves using the spectral element code SPECFEM3D with an idealized bathymetry across a $\SI{200}{\km}\times\SI{50}{\km}$ domain and resolving seismic waves up to 2.5~Hz, resulting in a \SI{560000} finite-element mesh.
The simulation ran for 8~h on 336 parallel processors to compute 200~s waveforms. 
\citet{Brissaud2017} coupled SPECFEM2D to a compressible Navier-Stokes solver to simulate seismo-gravito acoustic waves.
A 3D fully coupled elastic-acoustic simulation of the 2018 Palu, Sulawesi, earthquake and tsunami using the Discontinuous Galerkin code SeisSol and a mesh with 518~million elements (261~billion degrees of freedom) used 3072 nodes (147456 CPU cores) of the supercomputer SuperMUC-NG and took 5.5~h to compute \SI{30}{\s} of the event \citep{krenz20213d}. 
Seismically induced acoustic signals in air travel more slowly, thus requiring more refined meshes than ocean acoustic models to resolve the relevant frequencies.
\par
Modeling high-frequency seismo-acoustic wavefields in 3D therefore requires expensive numerical simulations using substantial computational resources on supercomputers because the number of required degrees of freedom typically scales with the minimum velocity and the highest resolved frequency \citep[e.g.][]{kaser_quantitative_2008}.
We use the open-source seismic wave propagation software SeisSol, which achieves a sustained high computational performance in the petascale range \citep{dumbserArbitraryHighorderDiscontinuous2006,uphoff_extreme_2017,krenz20213d,Ulrich2022}.
SeisSol has been extended to simulate coupled elastic-acoustic and tsunami gravity wave propagation \citep{krenz20213d,abrahams2022comparison}.
These advances allow us to numerically study the sensitivity of ground shaking and sound patterns to varying source properties.
\par
We compute high-resolution seismic and acoustic wave propagation scenarios across a $\SI{12}{\km} \times \SI{12}{\km}$ area with a sub-element refinement of up to \SI{2.3}{\meter}.
Our model is based on the largest \ML1.8 reference earthquake, event 13 in \citet{hillers_geothermal_2020}, induced during the Helsinki geothermal stimulation. 
We design and validate our simulations using the observed earthquake properties, a local velocity model, seismic ground motion measurements, acoustic recordings, and macroseismic reports (\cref{fig:overview}).
\par
In Section~\ref{sec:stimulation}, we summarize the Helsinki stimulation experiment and associated macroseismic reports that motivated our study.
In Section~\ref{sec:numexp}, we explain our multi-physics 3D model for fully coupled elastic-acoustic wave propagation scenarios in the Helsinki metropolitan area, which incorporates high-resolution topography data. 
In Section~\ref{sec:results}, we verify the numerical solution with seismic data and acoustic measurements.
We evaluate the effect of the source mechanism on the computed shaking and sound distributions which can be important data for modeling nuisance maps, as well as the consistency of the spatial variability of synthetic shaking and sound excitation with macroseismic observations.
Our analysis separates effects associated with $P$~wave and $S$~wave energy to assess the relative contribution of different body wave phases to sound disturbances.
This allows us to evaluate the compatibility of common assumptions and simplifications when approximating noise disturbance levels from seismic ground velocities.
The implications and limitations of our results are discussed in Section~\ref{sec:discussion}.

\section{The Stimulation Experiment, Data, and Macroseismic Reports}
\label{sec:stimulation}
%
In this section, we summarize the relevant data and observations associated with the Helsinki stimulation experiment that form the background for our study.
We study the seismic and acoustic wave excitations in response to two stimulations intended to establish a geothermal doublet system for district heating around \SI{6}{\km} below the Aalto University campus in Otaniemi, a district in the city of Espoo next to Helsinki (\cref{fig:overview}).
The majority of the induced seismicity is located below the shallow, $\sim$\SI{10}{\m} deep, and $\sim$\SI{3}{\km} wide Laajalahti Bay.
During the first larger stimulation in June and July 2018, approximately \SI{18000}{m^3} of freshwater was pumped into the deep crystalline rock formation, which induced thousands of small earthquakes \citep{kwiatek2019controlling}.
During the second smaller counter-stimulation in May 2020, a total volume of \SI{2900}{m^3} of water was used, and the seismic response was weaker \citep{Kwiatek2022}.
In 2018, seismicity was organized into three clusters between \SI{5}{\km} and \SI{6.5}{\km} depth \citep{kwiatek2019controlling}. A small cluster formed in response to the 2020 stimulation alongside the largest and deepest 2018 cluster \citep{Kwiatek2022}.
The $600{\times}1200$~m$^2$ lateral patch size of the seismicity around the two northeast-striking deep open-hole sections indicates that the source region is relatively compact compared to the area affected by the induced wavefields.
The local geology is characterized by Precambrian bedrock units.
Outcropping bedrock in backyards and open spaces is a common sight in the Helsinki area as it occupies approximately 25\% of the surface area.
The other parts are covered by meter-thick deposits of gravel, sand, clay, or peat.
Vegetated areas often thrive on a centimeter-thick topsoil layer covering the rocks.
At the topmost tens of meters, weathering processes have led to an approximately 50\% reduction compared with the high seismic basement rock velocities \citep{hillers_geothermal_2020}.
\par
The seismic response to the stimulations was monitored using a \SI{2}{\km} deep borehole string, 12 shallow borehole stations \citep{kwiatek2019controlling}, and a surface station network consisting of more than 100 sensors organized as single stations and mini arrays \citep{hillers_geothermal_2020,Rintamaki2021}.
Most stations were installed within a distance of \SI{20}{\km} from the well-head.
High sampling rates are essential for resolving the high-frequency energy of induced seismic wavefields that reach the surface.
Between 7 and 18 July 2018, the operating St1 Deep Heat Oy installed
two temporary microphone arrays, FIN1 and FIN2, at sites \SI{2.2}{\km} to the north-east and \SI{2.5}{\km} to the west from the well-head \citep{Lamb2021}.
The seismic data supports a number of studies focusing to date on
real time monitoring \citep{Ader2019},
reservoir and seismicity characterization \citep{kwiatek2019controlling,Bentz2020,leonhardtSeismicityStimulation1km2021,Kwiatek2022,Holmgren2023},
and wave-propagation effects \citep{Taylor2021,Li2021EGU}.
\citet{Eulenfeld2023} estimated high anelastic $Q$ values around $3{\times}10^5$ in the \SIrange[range-phrase={ to }]{10}{40}{\Hz} range and similarly high scattering $Q$ values around $1{\times}10^5$.
This exceptionally low attenuation explains the observed high transparency of seismo-acoustic waves at human-audible frequencies~\citep{Lamb2021}.
\par
A three-tier traffic-light system was defined.
For this, the probabilities to exceed peak ground velocity (PGV) and peak ground acceleration (PGA) thresholds
were linked to local earthquake magnitudes using ground motion prediction equations \citep{Ader2019}.
The green, amber, and red PGV thresholds are \SI{0.3}{\mm \per \s}, \SI{1}{\mm \per \s}, and \SI{7.5}{\mm \per \s}, respectively.
The thresholds are set to minimize cosmetic building damage in accordance with building codes. In addition, macroseismic effects on human perception are considered \citep{westawayQuantificationPotentialMacroseismic2014,bommerFrameworkGroundMotionModel2017}.
The amber and red maximum local magnitude \citep{Uski1996} limits were \ML1.2 and \ML2.1, respectively.
An adaptive injection protocol, partially based on earthquake physics \citep{galisInducedSeismicityProvides2017}, helped control the largest estimated magnitudes of \ML1.8 and \ML1.2 during the 2018 and 2020 stimulations \citep{Ader2019,kwiatek2019controlling,Kwiatek2022}.
\par
While no event exceeded the threshold magnitude, the public reported 220 and 111 individual observations of ground shaking and audible disturbances to the macroseismic questionnaire of the Institute of Seismology, University of Helsinki, in 2018 and 2020, respectively \citep{hillers_geothermal_2020,Rintamaki2021}.
In 2018, the number of responses scaled approximately exponentially with magnitude \citep{Ader2019}, 
that is, the distribution did not saturate towards the typically observed sigmoid function \citep{Schultz2021a}.
Scattered reports started to trickle in for magnitudes around \ML1.0, and a maximum of 83 reports was received after the largest \ML1.8 event.
The response patterns are modulated by the activity level, and during the 2020 stimulation, local COVID-19 mobility restrictions likely played a role in the comparatively increased reporting activity \citep{Rintamaki2021}.
Locations are available by street address,
which shows that a reporting center is located around the Munkkivuori neighborhood about \SI{3}{km} north-east of the injection site. 
Most other reports arrived from a north-south oriented $\SI{5}{\km}\times\SI{10}{\km}$ large area bordering the drill-site to the west.
\par
Humans can generally perceive sounds in the range of \SI{20}{\Hz} to \SI{20}{\kilo\Hz} \citep{fastl2006psychoacoustics}.
Sound at \SIrange[range-phrase={ to }]{20}{200}{\Hz} is called low-frequency sound,
and infrasound refers to the generally inaudible range below 20~Hz \citep[e.g.][]{moller2004hearing}.
Infrasound is excited by natural phenomena including
wind, thunder, volcanic activity, bolides, avalanches, large animals, and earthquakes, whereas artificial infrasound is mostly generated by powered industrial equipment \citep{muhlhans2017infrasound}.
However, the lower frequency threshold limiting human audible sound levels is not sharply defined at \SI{20}{Hz} but depends on the sound pressure. At sufficiently high sound pressure levels, humans can perceive infrasound not only as a hearing sensation, but also as vibrations felt in various parts of the body \citep{moller2004hearing},
which differs from the external vibrations of the ground or buildings caused by seismic waves \citep{Rutqvist2014perception}.
\par
In Helsinki, the simultaneous excitation of ground shaking and sound waves at frequencies around the lower limits of perceptible and audible frequency ranges caused a variety of sensations.
These include felt shaking and vibrations, 
infrasound effects that are difficult to interpret and that result in reports on mixed or combined sensations,
and audible effects that are typically described as blasts or thunder-like.
Residents navigated to the online macroseismic questionnaire maintained by the Institute of Seismology at the University of Helsinki.
Reports in response to a \ML1.2 event that occurred at 03:50 local time include the following translated statements.
\enquote{The whole family was awakened by an immense boom, similar to a fierce thunderstorm.}
\enquote{A loud boom was heard indoors.}
\enquote{A rock-based shaking. Lying on the bed it could be heard and felt as a vibration.}
This statement relates to a \ML1.0 event at 08:06 local time: \enquote{After the boom the movement propagated beneath the feet from the heels to the toes, in north or north-east direction.}
Quotes from responses to a \ML1.8 event at 20:35 include \enquote{The shaking was felt in the whole house, objects moved on the balcony} and
\enquote{A sharp metallic bang, no echo. Simultaneously, the apartment shook.}
\par
Audible nuisance generated by low-magnitude events prompted the developer to deploy microphone arrays.
Flanked by outreach efforts, the general attitude toward the stimulation did not critically decline \citep{Ader2019}, but some reports (\enquote{earlier on the same day and during previous days there were similar blasts that were really disturbing}) imply that weeks-long exposure to felt and heard disturbances has the potential to negatively impact public opinion on the development stages of geothermal energy production.
It has been speculated that resonance effects associated with the Laajalahti Bay are responsible for the sound excitation \citep{Ader2019}.
Thus, it seems plausible to expect more sound reports from around the Laajalahti Bay compared to the obtained distribution.
However, the resonance hypothesis was ruled out by comparing co-located borehole seismograms and microphone array data \citep{Lamb2021}, which showed that sound typically starts at the onset of the $P$~wave \citep{hill1976earthquake,Tosi2000}, which suggests local excitation.
This is compatible with the overall co-located distributions of felt and heard reports, that is, there are no areas with excess reports of audible noise in the absence of reported shaking, which would imply efficient sound propagation.
The array data analysis further suggests that later arriving acoustic waves propagating horizontally at the speed of sound are excited by secondary sources associated with the built environment, considering the absence of steep topographic interfaces.
The delay here is on the order of a few seconds \citep{Lamb2021}.
This implies that secondary sources were located within the epicentral area.
\par
\citet{hillers_geothermal_2020} compare the locations of the obtained reports to the
peak amplitudes at the surface of the $P$~wave, $SV$~wave, and $SH$~wave radiation patterns which are associated with the predominantly thrust faulting mechanisms of the induced seismicity.
Surface waves are not excited by the deep seismicity.
Visual inspection of radiation patterns shows good agreement between reporting activity, dominated by the Munkkivuori neighborhood, and the $SH$~amplitude distribution in that area.
However, the weaker correlation between the $SH$~radiation pattern and the locations of the responses to the largest \ML1.2 event during the 2020 stimulation questions a simple correlation between reporting activity and specific radiation patterns.
In this study, we compute the spatial distributions of peak ground velocities and peak sound pressures to analyze the seismo-acoustic coupling that is typically associated with the vertical ground motion of $P$~waves and $SV$~waves \citep{averbuchLongrangeAtmosphericInfrasound2020, Brissaud2017, Brissaud2021}.
Our numerical simulations explore the sensitivity of shaking and sound patterns to the properties of the wavefield and the earthquake source.
Focusing on scenarios associated with the largest induced \ML1.8 event in 2018, we address the computational challenges in modeling 3D sound propagation up to \SI{25}{\Hz}.
We show that numerical modeling helps to better understand the physical mechanisms that govern the public response to shaking and noise. Our approach can help mitigate seismo-acoustic nuisance associated with geothermal stimulation.
We discuss but do not quantitatively analyze socioeconomic effects.
%
\section{Numerical Experiments}
\label{sec:numexp}

\begin{table}[]
    \centering
\caption{Five investigated earthquake source mechanisms. The reference \textit{Event 13} is the largest \ML1.8 event induced during the stimulation that occurred on 16 July 2018 \cite[table 1]{hillers_geothermal_2020}. The next three source mechanisms are obtained from the reference event by rotating one angle by \SI{90}{\degree}. The \textit{Orthogonal} solution has a slip vector that is orthogonal to the first two mechanisms.}
\label{tab:earthquake catalogue} 
    \begin{tabular}{l*{5}{|r}}
        Event  & {\textit{Event 13}} & \textit{Strike+90} & \textit{Dip+90}   & \textit{Rake+90}   & \textit{Orthogonal}  \\
        \hline
        Strike (\si{\degree}) &  328  & 58   & 148 & 328 &  216 \\
        Dip (\si{\degree})    &   31  & 31   &  59 &  31 &   52 \\
        Rake (\si{\degree})   &   71  & 71   & 289 & 161 &   91 \\
        Beachball             & \includegraphics[width=1cm]{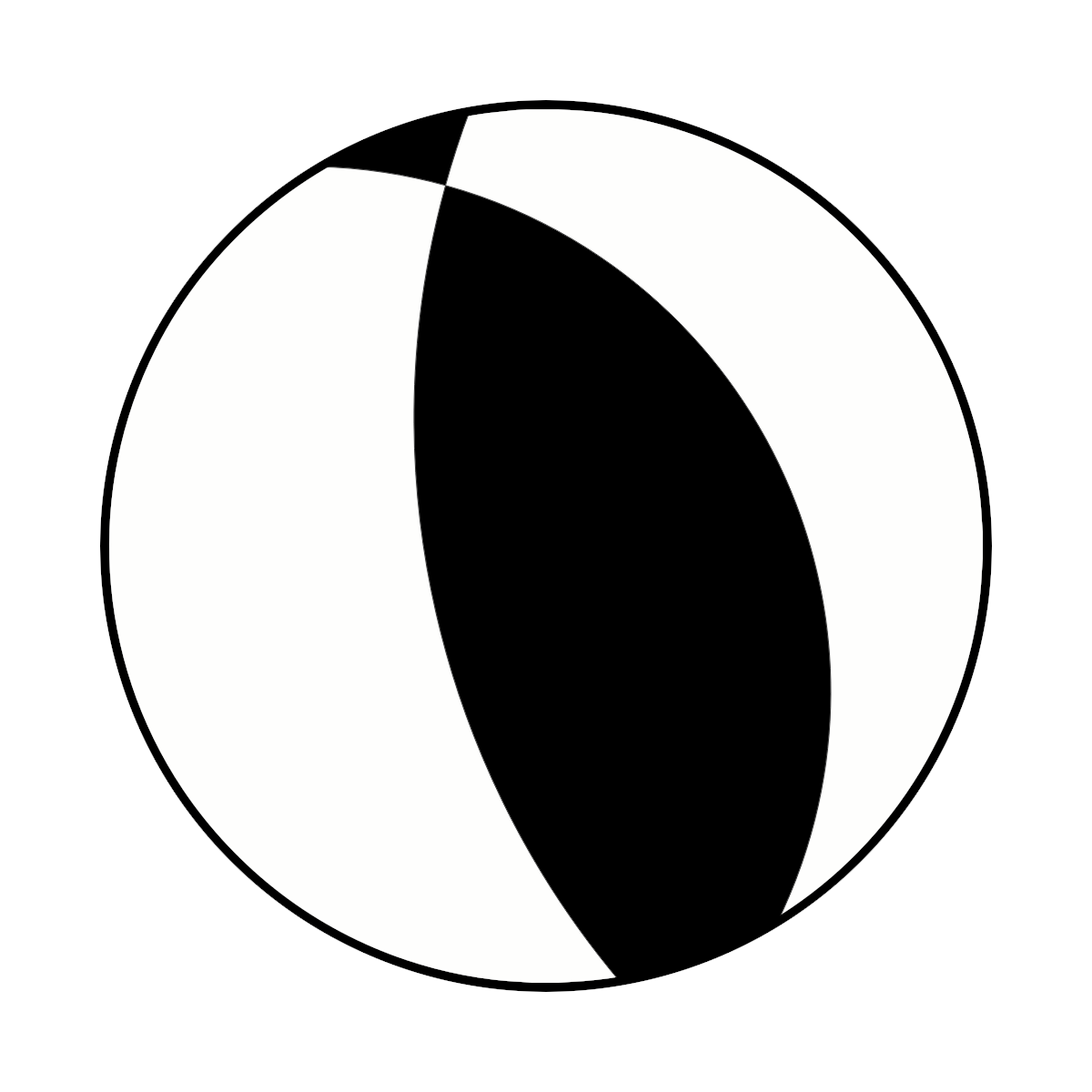} & \includegraphics[width=1cm]{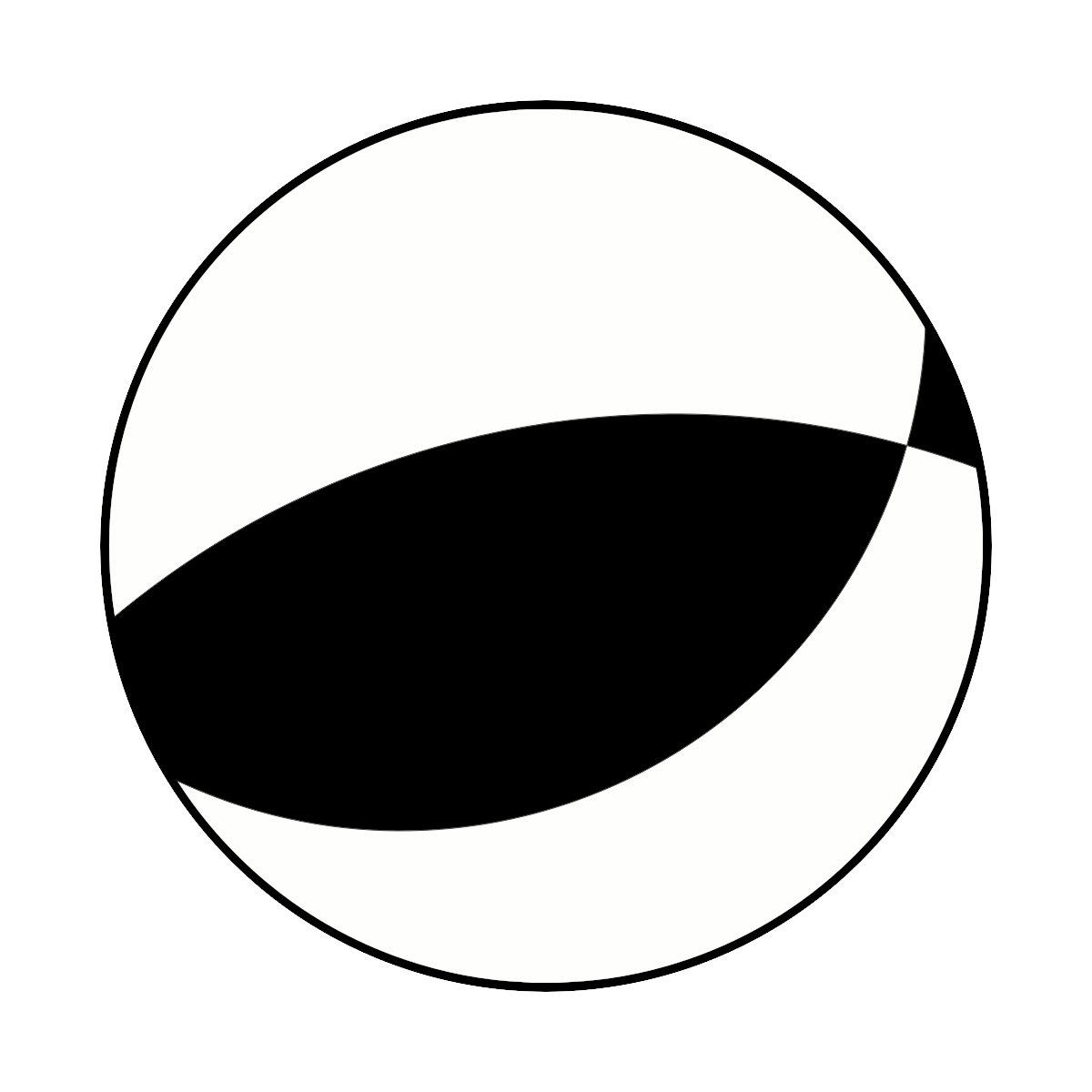} & \includegraphics[width=1cm]{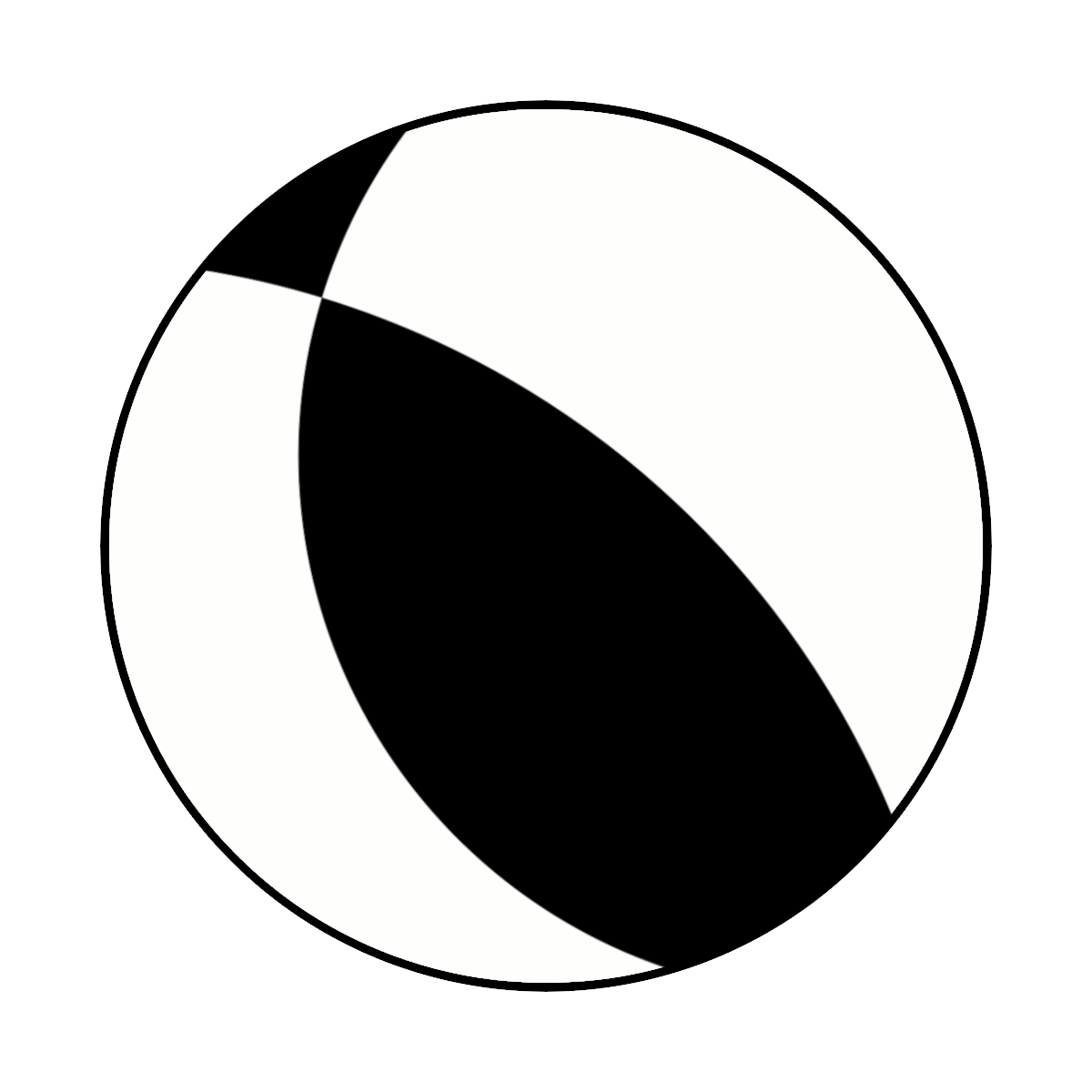} & \includegraphics[width=1cm]{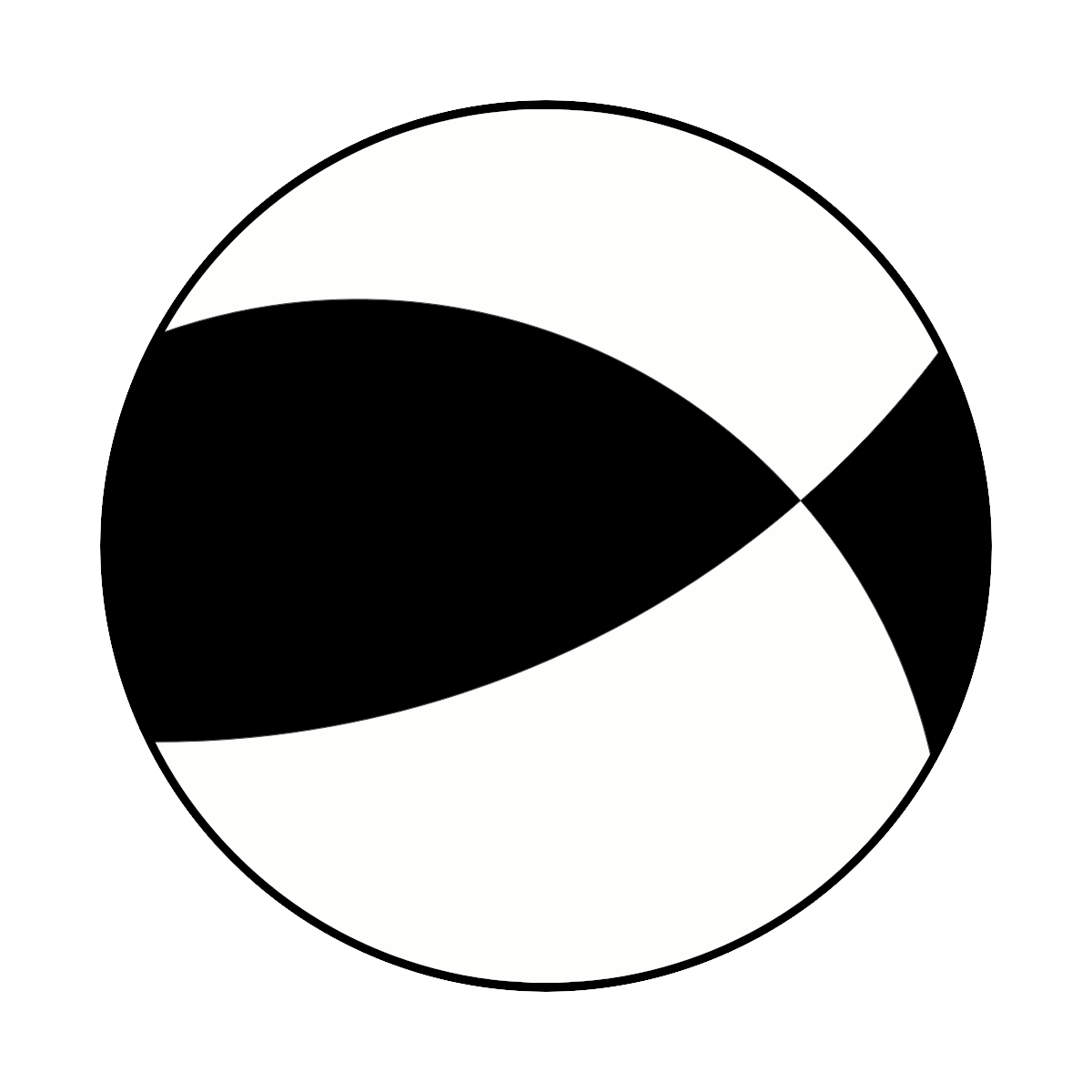} & \includegraphics[width=1cm]{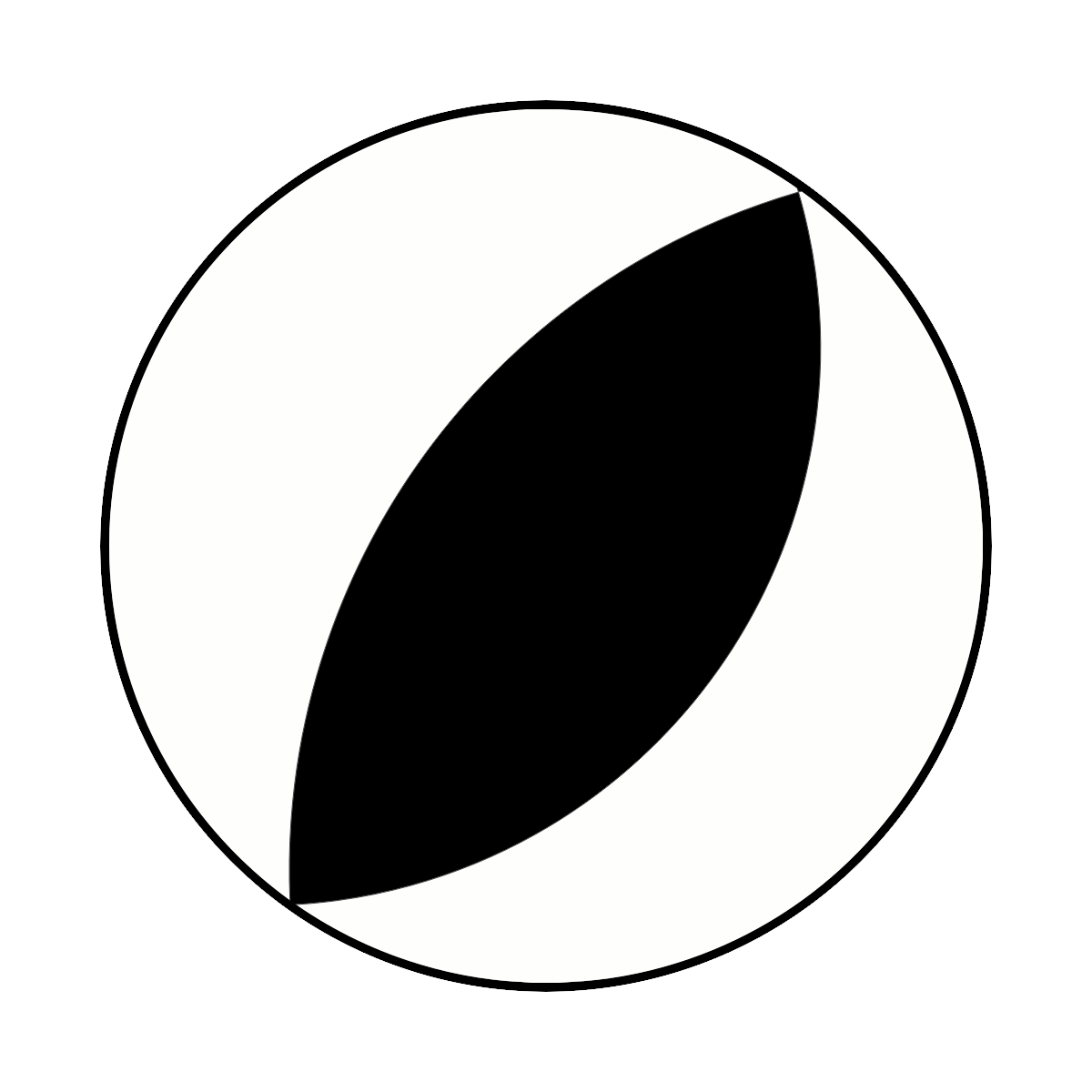} \\
    \end{tabular}

\end{table}

\subsection{Earthquake Source}
We parameterize earthquake point sources at the hypocenter location of the largest \ML1.8~event that occurred on 16~July~2018 at 17:26:03~UTC.
This earthquake was located at \SI{60.196}{\degree}N, \SI{24.837}{\degree}E at a depth of \SI{6.1}{\km}.
The source mechanism and macroseismic response to this event are detailed in~\cite{hillers_geothermal_2020}.
We convert the local magnitude to the seismic moment following \citet{saari2015evaluating,kwiatek2019controlling} as
\begin{equation}
  M_0 = 10^{(M_L + 7.98)/0.83}.
\end{equation}
We then use a Brune source time function
\begin{equation}
S(t) = \begin{cases}
\exp \left(-(t - t_0)/T \right) (t - t_0) / T^2 & (t - t_0) > 0 \\
0 & \text{else}
\end{cases}
\end{equation}
where $t_0{=}0.05$~s controls the onset time of seismic moment release, and $T{=}0.02$~s governs the source duration \citep{bruneTectonicStressSpectra1970,madariagaEarthquakeScalingLaws2011}.
This results in a seismic point source with a corner frequency of approximately \SI{24}{\Hz}.
Specifically, our chosen source time function yields a seismic moment amplitude at \SI{24}{\Hz}, which is a factor of ten smaller than the peak amplitude over the entire frequency range.
It is difficult to constrain the kinematics of small earthquakes observationally \citep{abercrombie2021resolution}, but relatively short rupture durations in the range of our choice of $T{=}\SI{0.02}{\s}$ have been reported \citep[e.g.,][]{tomic2009source}.
We analyze the effects of source geometry on the seismo-acoustics by using four additional scenario earthquake sources that are obtained by rotating the original moment tensor
(\Cref{tab:earthquake catalogue}).

\subsection{Velocity Model}
Our velocity model includes the solid Earth coupled with a \SI{2}{\km} thick air or atmospheric layer.
Below the ground surface, we use the 1D seismic velocity model of~\citet{leonhardtSeismicityStimulation1km2021} obtained by local vertical seismic profiling.
In this model, the velocity increases from $\sim$\SI{5.9}{\km \per \s} at the surface to $\sim$\SI{6.5}{\km \per \s} at \SI{3}{\km} depth, and then decreases towards \SI{6}{\km \per \s} at \SI{6}{\km} depth.
We compute the $S$~wave speed $v_S$ from the available $P$~wave speed $v_P$ using an empirical constant $v_P/v_S{=}1.71$ ratio \citep{leonhardtSeismicityStimulation1km2021}.
We assume a constant density of $\SI{2700}{\kg \per \m^3}$ for the entire domain.
For the air layer, we use a constant sound speed $c{=}\SI{340.5}{\m \per \s}$ and a constant density, $\rho{=}\SI{1.225}{\kg \per \m^3}$.

\subsection{The Computational Mesh}
%
%
%
%
%

We simulate 3D elastic wave propagation in a domain of $\SI{12}{\km} \times \SI{12}{\km} \times \SI{15}{\km}$ centered around the source location.
The coupled 3D acoustic wave propagation is modeled in the \SI{2}{\km} thick air layer (Fig.~\ref{fig:volume}).
At the interface between the elastic and acoustic subdomains, we consider accurate topography data from the National Land Survey of Finland with a resolution of \SI{2}{\m} and interpolate it to the unstructured triangular grid of our surface mesh.
The combination of slow sound speed in air and the need to resolve audible frequencies poses a significant computational challenge.
To keep the resulting computational cost manageable, we limit the highest resolution area to a cone-shaped refinement region (\cref{fig:overview}c).
This region includes both the source of the earthquake and the Munkkivuori neighborhood, from where most complaints about sounds were submitted. 
Inside this region, we set the element sizes to \SI{97}{\m} and \SI{14}{\m} for Earth and air, respectively.
To enforce a conforming mesh at the elastic-acoustic interface, our mesh contains elastic elements that are nearly as small as the acoustic elements.
The topography data further complicate the meshing process and can lead to element sizes much smaller than \SI{14}{\m} in both the elastic and acoustic domains.
For example, \SI{1}{\percent} of all element edges are smaller than \SI{7.04}{m}.
Outside this refinement region, we gradually decrease the mesh resolution to a maximum mesh element size of \SI{2}{\km}. The largest elements serve as cost-effective sponge layers, preventing artificial reflections from imperfect absorbing boundary conditions.
Our simulations do not conserve energy because energy is allowed to leave the domain at the absorbing boundaries.
\par
Within each element, we approximate the solution spatially with a fifth-degree polynomial, leading to sixth-order accuracy in space and time \citep{kaser_quantitative_2008}. This sub-element resolution allows us to achieve an effective resolution of \SI{2.3}{\m} in air and \SI{16.2}{\m} in Earth within the high-resolution region of interest. 
The resulting mesh contains $40.9$~million elements, of which only $2.6$~million correspond to our Earth model. The vast majority of the computational cost stems from the modelling of acoustic wave propagation in air.
By employing polynomial basis functions of fifth degree, we obtain \num{504} degrees of freedom per element.
Thus, our mesh includes \num{20.6}~billion degrees of freedom.
For comparison, we created a computationally cheaper setup with a uniform mesh resolution of \SI{70}{\m} in the Earth.
This setup does not model wave propagation in air and contains 32.5~million elements.
Including an air layer with such a high resolution for the entire domain is extremely challenging with currently available computational infrastructure. We estimate that such a mesh would contain more than \num{500}~million elements and thus more than \num{250}~billion degrees of freedom.


\subsection{Cluster-Based Local-Time-Stepping Algorithm}
The mesh refinement in the high-resolution region, together with the stark contrast in element size between the elastic and acoustic layers, leads to massive differences in element sizes.
Because SeisSol uses an explicit time-integration method for its wave propagation solver, the standard Courant–Friedrichs–Lewy time step restriction \citep{courantUeberPartiellenDifferenzengleichungen1928} leads to strongly diverging permissible time step sizes across the elements.
To tackle such situations, SeisSol offers a cluster-oriented local-time-stepping (LTS) method that groups elements into clusters, according to their admissible time step size \citep{breuer_petascale_2016,uphoff_extreme_2017}.
Each cluster is updated in a multi-rate fashion, as often as needed locally.
This LTS scheme significantly reduces the time-to-solution of our scenarios and is therefore crucial for the feasibility of our simulations.
In our fully coupled scenarios, the minimum time step is \SI{8.4e-06}{\s}. 
The maximum time step size is 2048 times larger.
Our LTS strategy yields an approximate speedup factor of 30 for the fully coupled seismo-acoustic model.

\subsection{Output and Postprocessing}
\label{subsec:output}
\subsubsection{Simulated Output}
\begin{figure}
    \centering
    \includegraphics[width=\textwidth]{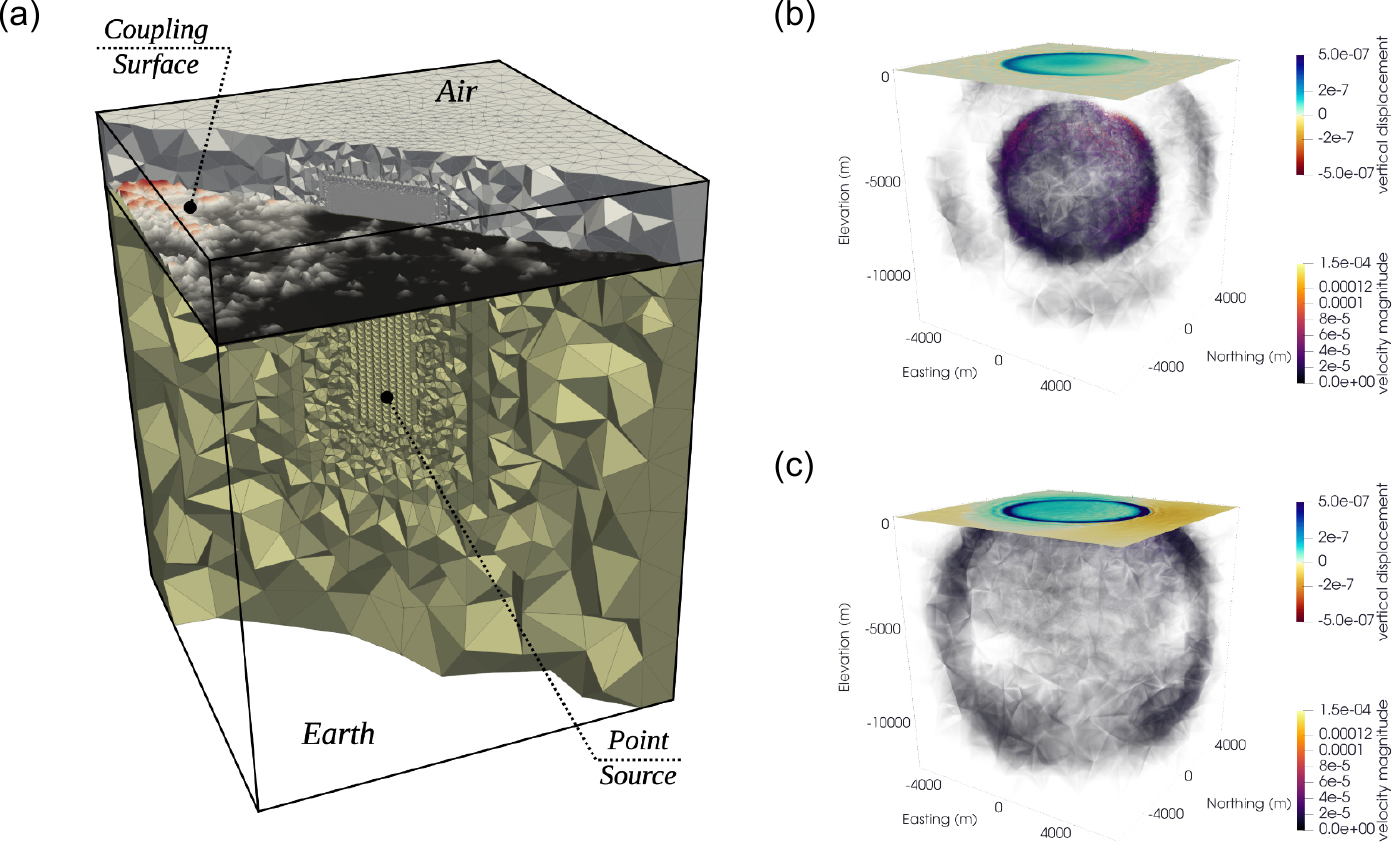}
    \caption{
    (a) View of our computational mesh with the elastic layer in the lower part and the thinner acoustic layer on top that contains the refined mesh in the central region. We highlight the topography at the interface.
    (b), (c) Velocity field in the volume and vertical displacement at the surface at (b) \SI{1.2}{\second} and (c) \SI{2.0}{\second} for the simulations of event 13, the largest induced earthquake. 
    The $x$-axis points in the west-east direction, the $y$-axis in the north-south direction, and the $z$-axis up-down. The view is from the south-east towards north-west. Note the larger vertical displacement value associated with the $S$~wave in (c).}
   \label{fig:volume} 
\end{figure}

\begin{figure}
    \centering
    \includegraphics[width=\textwidth]{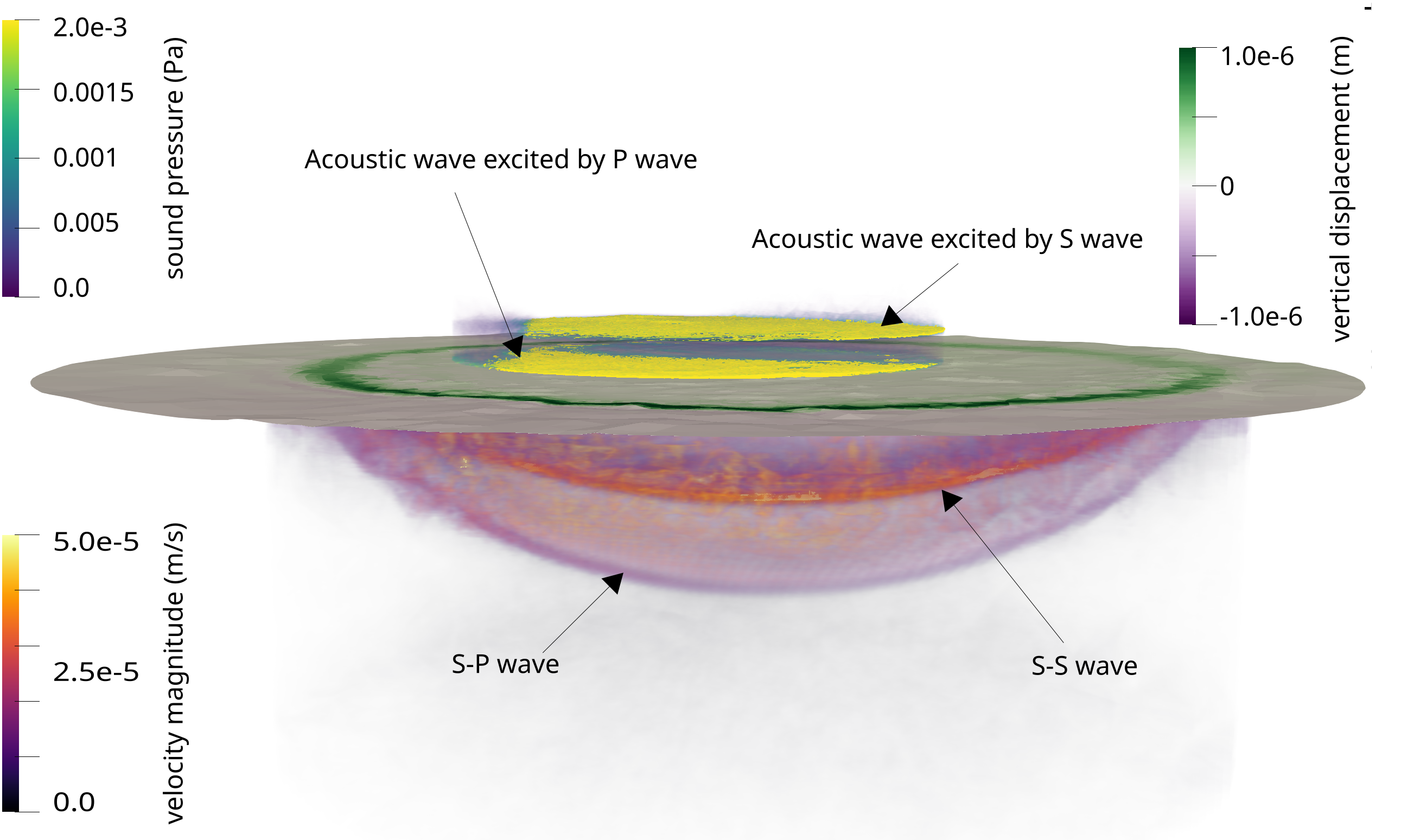}
    \caption{\label{fig:volume-ufo}%
    Acoustic and elastic wavefields at $t{=}\SI{2}{s}$. The acoustic wavefield is illustrated in a cylinder with \SI{2}{\km} radius around the source and the seismic wavefield is shown in a \SI{4}{\km} radius. The illustration features two wave fronts in the acoustic layer, the upper one excited by the $P$~wave and the lower one excited by the $S$~wave. The elastic layer contains the reflected $S$~wave and the reflected $P$~wave that are both generated by the interaction of the incident $S$~wave with the elastic-acoustic interface.}
   
\end{figure}

\begin{figure}
    \centering
    \includegraphics[width=\textwidth]{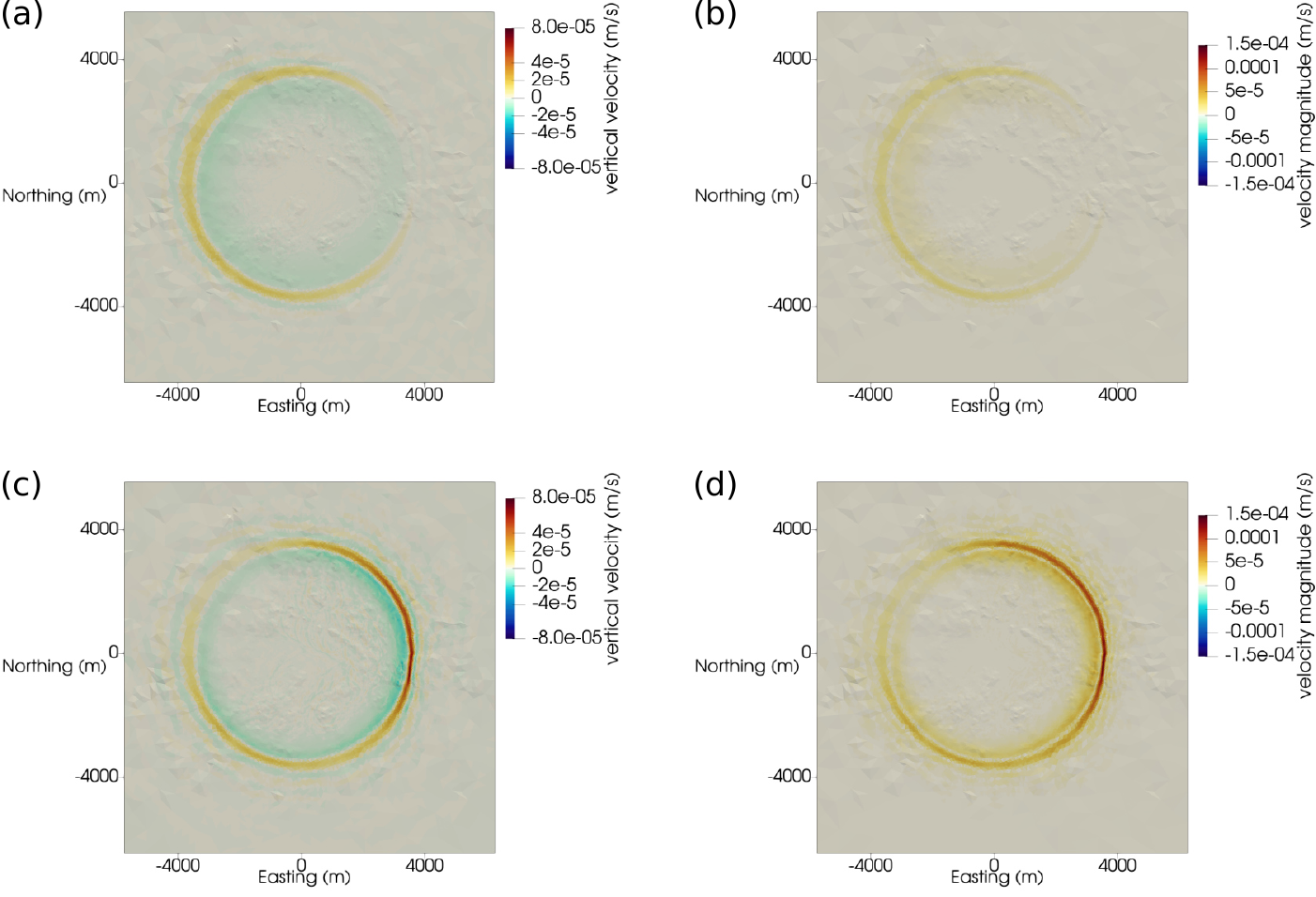}
    \caption{
    (a), (b) Vertical ground velocity and velocity magnitude at \SI{1.2}{\second} for the \ML1.8 event 13 at the Earth surface.
    The timing is associated with the $P$~wave.
    (c), (d) Vertical ground velocity and velocity magnitude at \SI{2.0}{\s}. The timing is associated with the $S$~wave.
    }
    \label{fig:free-surface}
\end{figure}

In the fully coupled simulations (\cref{fig:volume-ufo}), we place a grid of receivers within our high-resolution model area with a spacing of \SI{100}{\m} at an elevation of \SI{0.5}{\m} above the free-surface to record the synthetic acoustic fields.
The seismic wavefield is sampled at receivers located just below the surface at a depth of \SI{0.05}{\m}.
The receiver output is written at a sampling rate of 200 times per second.
For the Earth-only simulation, we write the velocity and displacement fields at the entire free surface at a rate of 1000 times per second.
For a pointwise comparison of recorded data and simulated synthetics, we add receivers at the locations of the four FIN2 acoustic stations (inverted triangle in \cref{fig:overview}) from \cite{Lamb2021} at the \SI{0.5}{\m} elevation.
We place approximately 100 seismic receivers at all the locations of the ST1 borehole sensors and surface stations of the 2018 HE and OT monitoring networks \citep{hillers_geothermal_2020}.
The triangles in \cref{fig:overview} represent the subset used for data comparison.
For both the Earth domain-only and coupled simulations, we compute the released energy perturbation at a rate of ten times per second.
We visualize the vertical velocity and the velocity magnitude, that is, the length of the velocity vector, recorded at the Earth's surface in \cref{fig:free-surface}.
Superimposed on the regional topography with maximum elevations on the \SI{10}{\m} scale,
these illustrations show source effects, such as the radiation patterns of the $P$~wave and $S$~wave. We observe the expected smaller $P$~wave amplitudes and path effects associated mainly with topography scattering.
Although scattering effects from the topography do not cause strong decoherence of the near-source body wavefronts, they do cause visible coda arrivals \citep[e.g.,][]{Pitarka2022,Taufiqurrahman2022}.
We additionally visualize the 3D volumetric fully coupled wavefields at selected time steps.
In \cref{fig:volume}, we show the magnitude of the velocity wavefield in the volume and vertical displacement at \SI{1.2}{\second} and at \SI{2.0}{\second}.
\Cref{fig:volume}b illustrates the four lobes of the $P$~wave, which have partially reached the Earth's surface. In the same instance, stronger $S$~wave amplitudes are shown closer to the source.
In the later snapshot in \cref{fig:volume}c, the $P$~wave energy has propagated beyond the computational domain, and the $S$~waves start to interact with the Earth's surface.
In \cref{fig:volume-ufo}, we illustrate the interaction of the elastic and the acoustic waves at the interface.
At \SI{2}{\s} the seismic $P$~wave has propagated out of the computational domain, and we only see the reflected $S$~wave in the solid Earth.
In the atmosphere, we can distinguish two wave fronts that are consecutively excited by the seismic $P$~wave and the $S$~wave.
Hence, although the seismic $P$~wave is outside the domain, 
the corresponding acoustic wave is still propagating inside the computational domain due to the comparatively low sound speed.

\subsubsection{Calibration}
A commonly used relationship that relates the pressure perturbation $\Delta P$ to the vertical ground velocity $v$ is
\begin{equation}\label{eq:pressure-rule-of-thumb}
    \Delta P = \rho c v,
\end{equation}
where $\rho$ is the density of air and $c$ the speed of sound \citep{cookInfrasoundRadiatedMontana1971,Tosi2000,Lamb2021}.
This result can be obtained by assuming a vertically incident plane wave, which is an exact solution to the acoustic wave equation.
Our 3D simulations including topography provide the opportunity to evaluate this approach.
For each pair of the 1386 densely spaced elastic and acoustic receivers, we compute the peak ground velocity of the elastic part and the peak pressure perturbation of the acoustic part.
We perform a linear regression assuming the following relation
\begin{equation}
\label{eq:linear-regression}
   \Delta P^\text{peak} = c_0 + c_1 v^\text{peak} + \varepsilon,
\end{equation}
where $\varepsilon$ is normally distributed error term.
The constant factor $c_0$ is zero if the vertical plane wave assumption holds.
It can be invalidated for reasons such as topography or due to source effects.
Here, however, it captures the average difference between the perfectly linear model, given by \cref{eq:pressure-rule-of-thumb}, and the measurements.
Similarly, we would expect that $c_1{=}\rho c {\approx} \SI{417.1}{\kg \per (\m^2 s)}$.
\par
We implement a workflow that benefits from the combination of our two setups together with this calibrated rule of thumb.
First, we simulate the scenario that includes Earth and air.
From the refinement zone, we extract the peak ground vertical velocity in the Earth and the sound pressure level in the air.
We perform the linear regression \cref{eq:linear-regression} to obtain the approximate relationship between these two quantities.
The result can be considered as a calibrated version of \cref{eq:pressure-rule-of-thumb} that takes into account factors such as topography and vertical and horizontal distance to the source.
Next, we compute the refined Earth model without the acoustic coupling, allowing for computational efficiency, and extract the peak vertical ground velocity for the entire domain.
Finally, we use the result from the linear regression to estimate the sound pressure level.
This combines the strengths of the 3D fully coupled approach with the computational efficiency of the Earth-only approach.

\subsubsection{Phase Estimation}
We investigate whether the $P$~wave or the $S$~wave has a larger impact on the disturbance patterns.
We estimate the arrival time $t^S$ of the $S$ wave using the simple relation $t^S(d){=}d/c^S_\text{max}$, where $d$ is the distance to the source and $c^S_\text{max}{=}\SI{3.83}{\km \per \s}$ is the maximal $S$~wave speed in our velocity model.
We use this to define the arrival windows.
A point with hypocentral distance $d$ is assumed to be affected by the $P$~wave at times $t \in (0, c^S_\text{max})$ and by the $S$~wave for all other times.
As we use a lower bound for the $S$~wave arrival time, we may underestimate the duration of the $P$~wave.
However, this does not significantly bias our results because the $P$~coda amplitudes are smaller compared to the direct wave amplitudes.

\subsection{Computational Aspects}
We ran our setups for \SI{3}{\s} simulated time on the clusters SuperMUC-NG and Mahti.
On SuperMUC-NG, our simulations run on 200 nodes with 48 cores per node.
The fully coupled model takes about 1.25 hours for one simulation.
Simulations on Mathi run with similar efficiency.
For a more detailed analysis of the computational efficiency of large-scale fully coupled seismo-acoustic simulations with SeisSol, including the local-time-stepping method, we refer to \citet{krenz20213d}.

\section{Results}\label{sec:results}


\subsection{Comparison of Seismic and Acoustic Observations and Synthetics}
\begin{figure}
    \centering
    \includegraphics{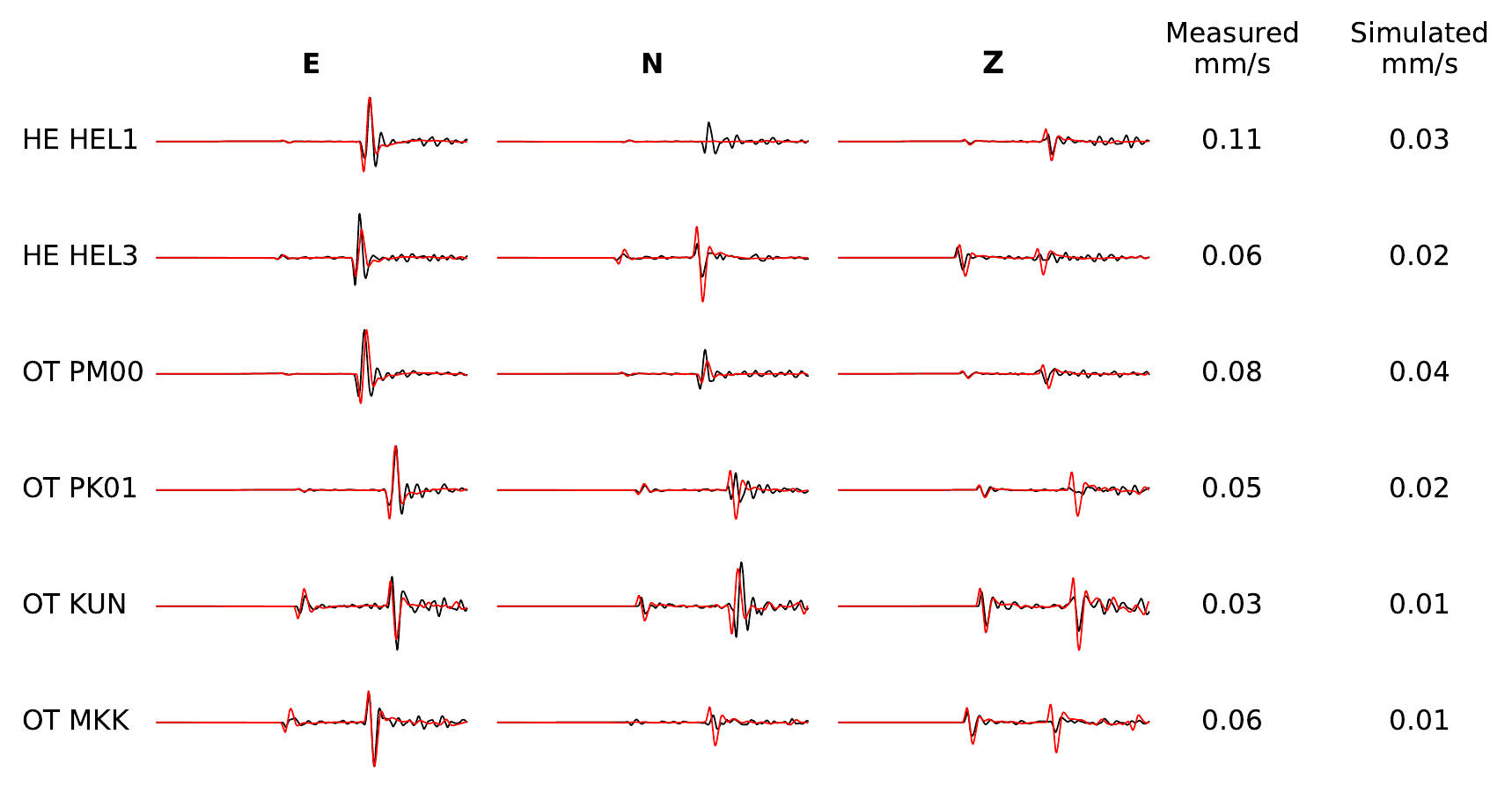}
    \caption{Comparison of observed (black) and synthetic (red) velocity waveforms at two broadband and four short-period seismic stations.
    The HE network stations are permanent broadband sensors, and the four OT network stations are temporary installations of \SI{4.5}{\Hz} geophones connected to CUBE recorders \citep{hillers_geothermal_2020}.
    We removed the instrument response from the seismic records using pre-filter corner frequencies of 0.5 and \SI{40}{\Hz} considering the bandpass filter range between 1 and \SI{10}{\Hz} applied to all data and synthetics.
    Synthetic and observed waveforms are aligned with respect to the $P$~wave arrival to account for the arbitrarily chosen onset timing of our source time function.
    This time shift can vary by \SI{0.02}{\s} between stations. It accounts for velocity heterogeneities that are not included in our 1D velocity model. 
    Synthetic and observed velocity waveforms are normalized individually by the maximum value of the records at each station, i.e., we do not normalize waveforms component wise.
    The peak velocity is indicated in the last two columns.
    }
    \label{fig:result-elastic-receiver-comparison}
\end{figure}

%
%
We begin the assessment of the synthetics with the comparison of \SIrange[range-phrase={ to }]{1}{10}{\Hz} filtered three-component seismograms.
Fig.~\ref{fig:result-elastic-receiver-comparison} shows waveforms of the reference \ML1.8 event at two broadband and four short-period stations located over a wide azimuthal range in the computational domain (triangles in \cref{fig:overview}).
Together with the relatively low frequency range, this selection allows us to focus on first order features, including $P$ and $S$~wave travel time, polarity, and  relative $P$ and $S$~wave amplitudes.
Visual comparison shows a relatively high consistency for all these characteristics between our synthetic (red) and observed (black) velocity waveforms. 
Considering the $P$~wave velocities between the surface and the injection depth in the range of \SIrange[range-phrase={ to }]{5.9}{6.5}{\km \per \s}, 
the overall high quality of the $P$ and $S$~waveform fits,
and the relatively small level of scattered waves below \SI{10}{\Hz}, the local geological situation in the southern Fennoscandian Shield on the here considered scales is well approximated by a homogeneous half-space model.
Correspondingly, the 1D velocity model obtained at the location of the injection \citep{leonhardtSeismicityStimulation1km2021} in the center of the domain is a good approximation for the 3D velocity structure in the area. 
The tens-of-meters thick low-velocity layer in the area inferred from an ambient noise surface wave dispersion analysis \citep{hillers_geothermal_2020} does not appear to be relevant for the body wave propagation below \SI{10}{\Hz}.
\par
The near-perfect consistency of the polarities, and the good agreement of the $P$-to-$S$~wave ratio between data and synthetics are governed by the source moment tensor properties that have been obtained from first motion polarity data \citep{hillers_geothermal_2020} and from waveform inversion \citep{Rintamaki2023EGU}.
The visible inconsistency between data and synthetics such as the $S$~wave properties at the N channel of the HEL1, PK01, KUN, and MKK stations suggest, again, a relatively small degree of structural heterogeneity,
which is compatible with wave propagation phenomena inferred by array analysis techniques \citep{Taylor2021,li2022rupture}.

%
%
\par
\begin{figure}[htb]
         \centering
         \includegraphics[scale=1]{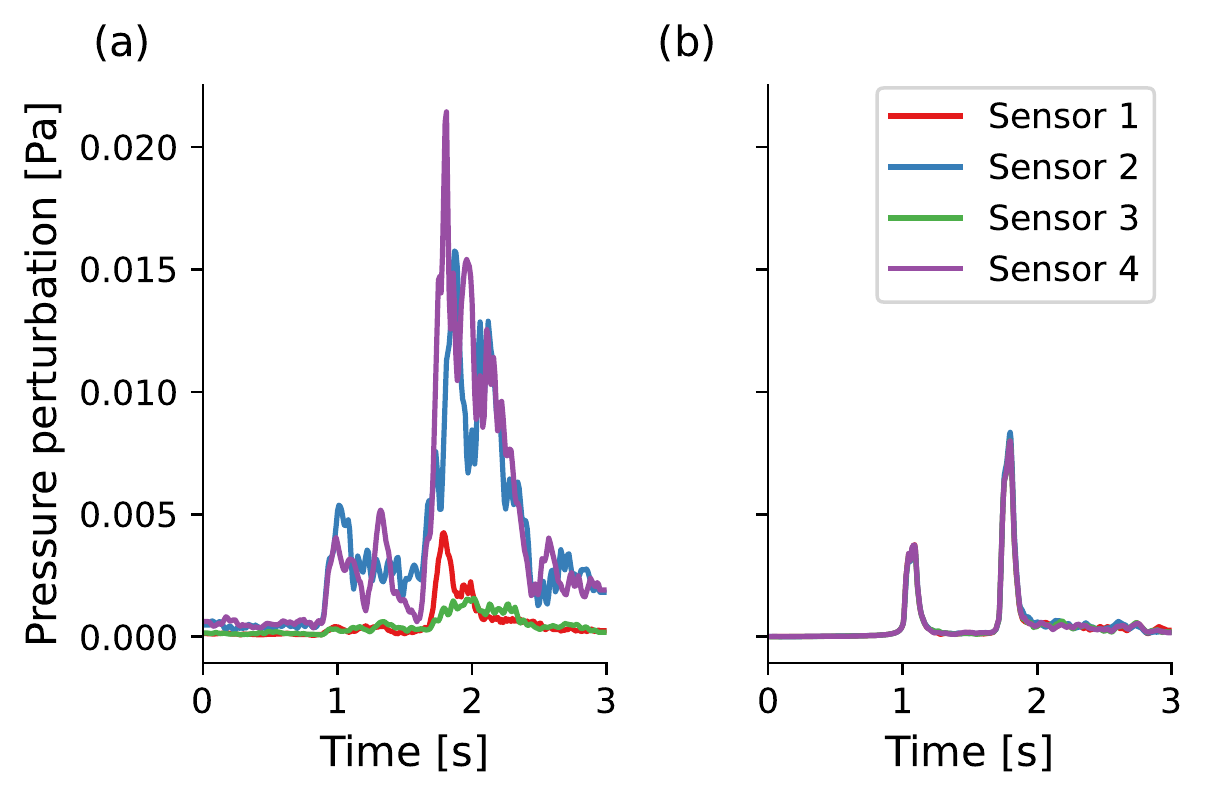}
         \caption{\label{fig:fin2-comparison-receiver}
         (a) Envelopes of acoustic FIN2 station data \citep{Lamb2021}. The inter-sensor distance is about \SI{10}{\m} to \SI{30}{\m}.
         (b) Envelopes of the synthetic acoustic simulation time series.
         For both observed and simulated data, we apply a \SI{1}{\Hz} high pass filter and smooth the envelope with a centered moving average filter with a window size of \SI{0.04}{\s}.
         Colors indicate different stations of the array.
         }
\end{figure}
\begin{figure}[htb]
         \centering
         \includegraphics[scale=1]{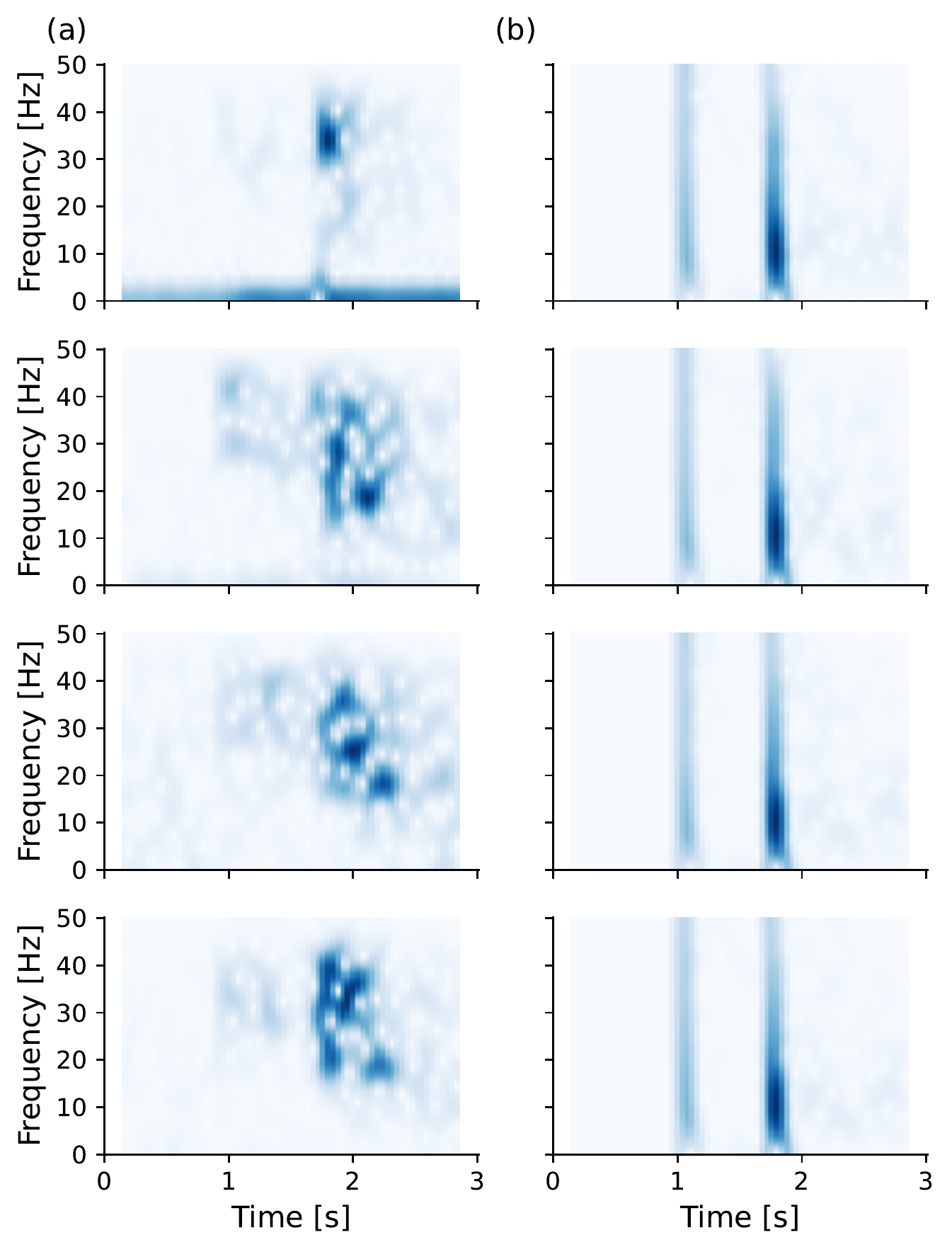}
         \caption{(a) Spectrograms of acoustic signals recorded at the four FIN2 stations computed with a window size of \SI{0.25}{\s}.
         (b) Synthetic spectrograms downsampled by factor of two to match the frequency range of the observed data. Accurate numerical resolution is limited to about \SI{25}{\Hz}, and synthetics at higher frequencies may be affected by numerical artifacts.
         Light and dark colors correspond to weak and strong squared power density, respectively.
         }
         \label{fig:fin2-comparison-spectogram}
\end{figure}

Similar to the seismogram validation, we also focus on general first order properties when comparing modeled acoustic time series to acoustic data recorded by the FIN2 microphone array \citep{Lamb2021}.
We neglect simultaneously obtained data from a FIN1 array \citep{Lamb2021} because of lower data quality.
To enhance the first order properties, we consider waveform envelopes (\cref{fig:fin2-comparison-receiver}) and spectrograms (\cref{fig:fin2-comparison-spectogram}).
These, too, show a good general consistency.
We highlight that the absolute amplitudes of acoustic energy are reproduced well to first order (\cref{fig:fin2-comparison-receiver}.
The envelopes are not normalized and represented at the same scale.
As for the seismograms, the relative arrival times of the $P$ and $S$~wave energy matches well.
Both data and synthetics feature higher energy associated with the $S$~wave arrival, which can be understood from the relative station position within the radiation pattern.
The sensitivity of acoustic data acquisition is demonstrated by the high variability of the four measured time series in \cref{fig:fin2-comparison-receiver}.
Considering the short direct $P$ and $S$ waveform segments that can be inferred from the synthetics in \cref{fig:fin2-comparison-receiver}, the variability between the four records is larger compared to the variability between the data average and the synthetics.
The strongest disagreement concerns the elevated sound pressure in the $P$~wave and $S$~wave coda that is insufficiently reproduced by the numerical simulations.
The acoustic coda can be related to secondary sources associated with the built environment
and to propagation effects that can include the scattering off turbulence and reflections off atmospheric boundary layers associated with temperature inversion.
This is compatible with wave propagation phenomena derived from array analysis techniques \citep[e.g.]{sminkThreeDimensionalArrayStudy2019}.
\par
The decreasing high-frequency content with time that can be discerned in the spectrograms in \cref{fig:fin2-comparison-spectogram} suggest that coda waves are likely controlled by scattered seismic waves that continue to excite sound locally. 
Compared to the envelopes, the acoustic spectrograms exhibit similar $S{-}P$ times and similarly stronger $S$~wave arrivals.
Importantly, however, observed acoustic energy and coda is excited at frequencies that exceed the here accurately resolved frequency range which is limited to approximately \SI{25}{\Hz}.
The high-frequency contributions in the synthetics above \SI{25}{\Hz} are likely affected by numerical dissipation and dispersion errors and noise, for example as a by-product of the employed high-order scheme that manifests as Gibbs phenomenon \citep{hesthaven_nodal_2008}.
Hence, despite the successful numerical synthesis of key seismo-acoustic wave propagation phenomena at the lower limit of the the human audible frequency range, 
the richness of the acoustic observations at higher frequencies are a reminder of the remaining challenges to model the full range of audible earthquake sounds.

%
%
%
\subsection{Linear Regression}
\begin{figure}[htb]
    \centering
    \includegraphics[scale=1]{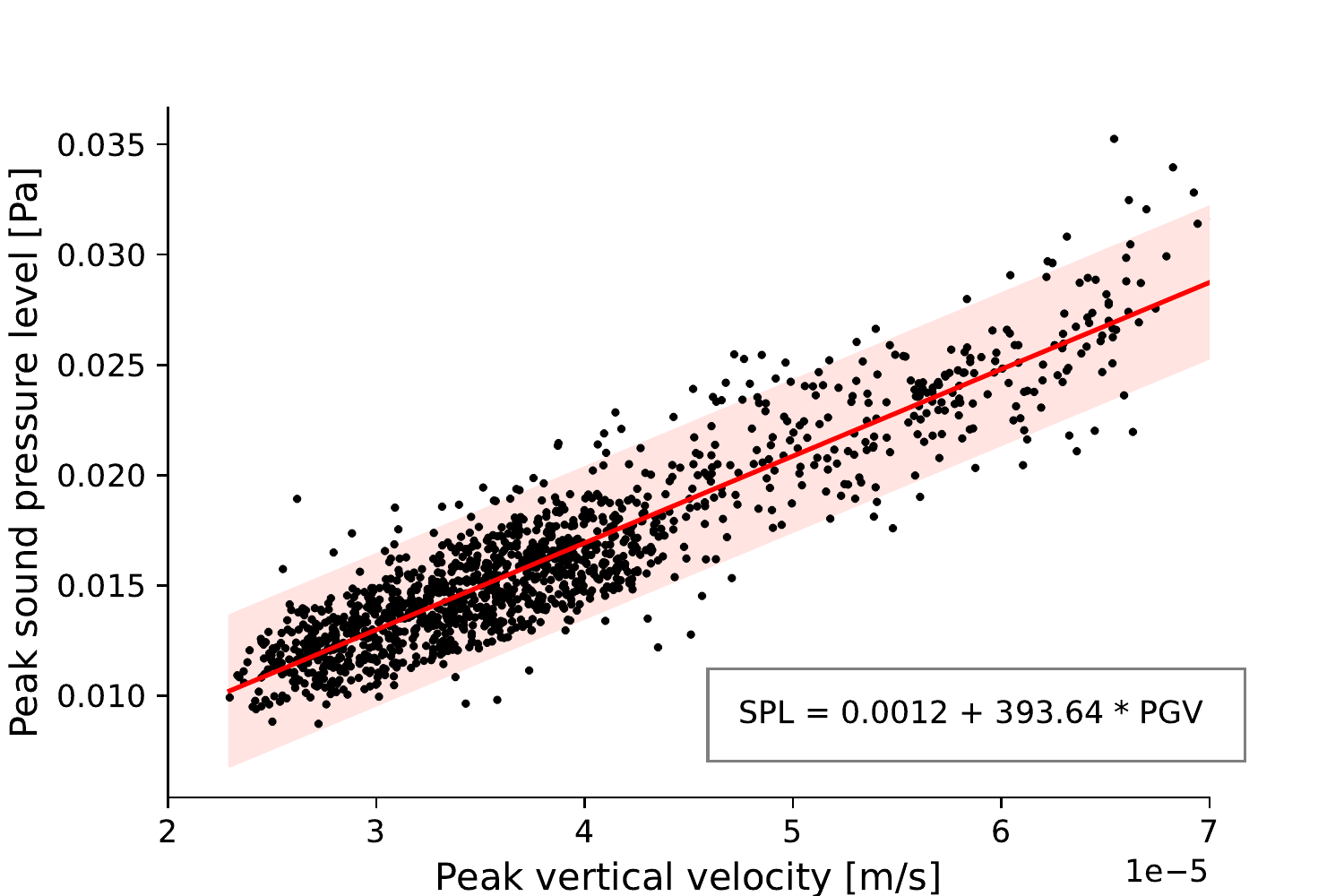}
    \caption{Relationship between the simulated peak ground velocity PGV and peak sound pressure level SPL inside the high-resolution model refinement zone for the reference \ML1.8 event 13.
    We computed this data with our coupled elastic-acoustic simulation.
    The red line indicates the linear regression fit, the shaded area is the 95\% prediction interval.}
    \label{fig:linear-regression-pgv-vs-spl}
\end{figure}

We compute the regression model for the five different source models individually using the statistical library \texttt{statsmodels}~\citep{seabold2010statsmodels}.
The results are compiled in \cref{tab:result-linear-regression}.
For all simulations, the intercept and slope are statistically significant with $p {<} 0.001$.
The intercept is non-zero for all considered simulations and the slope is smaller compared to the prediction \cref{eq:pressure-rule-of-thumb}.
However, the fit between the model and the synthetics associated with the reference event is good considering the obtained confidence interval and coefficient of determination.
With the exception of the Rake+90 scenario, 
the other considered rotations of the source moment tensor yield synthetics that show larger scatter around the model parametrization.
In detail, for the reference event, the linear regression results in values of $c_0 {=} 0.00118 {\pm} 0.00036$ and $c_1 {=} 393.64 {\pm} 9.11$, where the indicated uncertainty refers to the 95\% confidence interval.
This model has a coefficient of determination of $r^2{=}0.839$.
Compared to \cref{eq:pressure-rule-of-thumb}, which leads to a factor of $c_1{\approx}417$, our approximation results in a sound pressure level that is roughly 6\% smaller on average.
The data and regression results in \cref{fig:linear-regression-pgv-vs-spl} illustrate the high explanatory power of the linear relationship with only few outliers.
However, it is important to note that a significant number of data points with very high sound pressure level are not predicted well by the linear model.
We expect that an even higher resolution in the very near field to the epicenter, where high peak ground velocities are simulated, may help mitigate the remaining mismatches.
\begin{table}[htb]
    \centering
\caption{Linear regression results. The indicated uncertainty refers to the variation associated with a 95\% Student-t based confidence interval.
The quantity $r^2$ is the coefficient of determination.}
\label{tab:result-linear-regression}
 \begin{tabular}{
l
S[table-format=1.5]@{\,\( \pm \)\,}
S[table-format=1.5]
S[table-format=3.2]@{\,\( \pm \)\,}
S[table-format=2.2]
S[table-format=1.3]
}
\toprule
Source & \multicolumn{2}{c}{Intercept} & \multicolumn{2}{c}{Slope}
& { $r^2$ }     \\
\midrule   
Event 13 & 0.00118 & 0.00036 & 393.64 & 9.11 & 0.839 \\
Strike+90 & 0.00243 & 0.00048 & 356.22 & 13.34 & 0.665 \\
Dip+90 & 0.00155 & 0.00046 & 381.97 & 12.30 & 0.728 \\
Rake+90 & 0.00030 & 0.00019 & 414.62 & 7.97 & 0.883 \\
Orthogonal & 0.00265 & 0.00055 & 354.70 & 14.24 & 0.633 \\
\bottomrule
\end{tabular} 
\end{table}

%
%
\subsection{Peak Sound Pressure of $P$~Waves and $S$~Waves}
\begin{figure}
    \centering
    \includegraphics{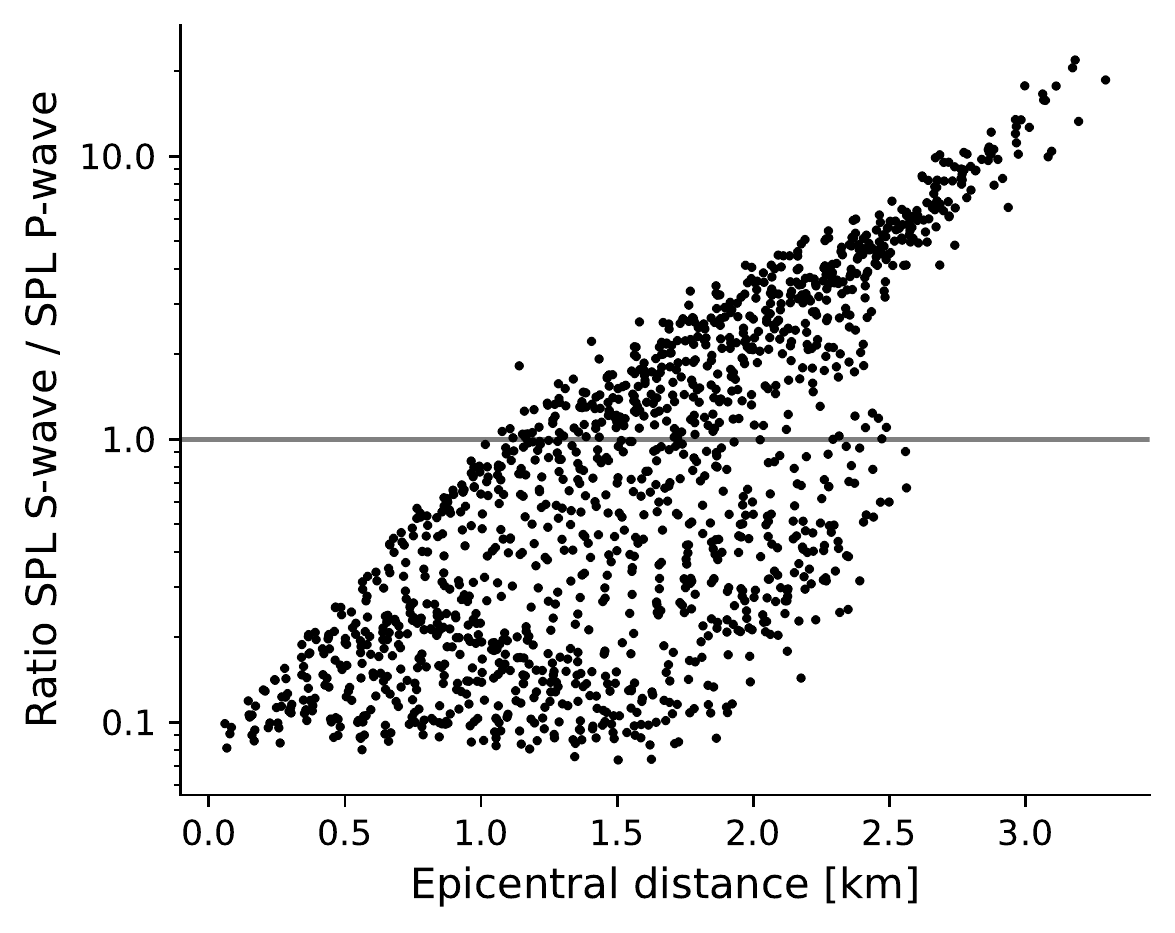}
    \caption{%
    Dependence of the simulated peak sound pressure level ratio on distance.
    We computed this data with our coupled elastic-acoustic simulation.
    The $x$-axis indicates the distance to the epicenter. The $y$-axis shows the peak sound pressure level SPL during the passage of $S$~waves normalized by the sound pressure level during the $P$~wave passage.
    The horizontal line at unity indicates balanced SPLs. Values above the line indicate excess SPL of the $S$~waves.
    The values are obtained from the fully coupled simulation for the reference \ML1.8 event 13 earthquake.}
    \label{fig:result-spl-s-p-ratio}
\end{figure}
We evaluate the correspondence between the $P$~wave and $S$~wave and the peak sound level computed by our simulation in the high-resolution area.
The $P$~phase is commonly assumed responsible for earthquake sound generation~\citep{hill1976earthquake}.
To verify this, we use the previously defined $P$~wave and $S$~wave windows (\cref{subsec:output}).
Using the reference event synthetics,
\cref{fig:result-spl-s-p-ratio} shows that the $P$~wave dominates the sound generation at short epicentral distances.
The relative contribution of acoustic signals generated by the $S$~wave becomes increasingly stronger with distance, and at about \SI{1}{\km} epicentral distance the peak sound pressure is associated with the impinging $S$~wave.
\subsection{Peak Ground Velocity and Sound Pressure Level Distributions}
%
%

\begin{figure}
    \centering
    \includegraphics[scale=0.9]{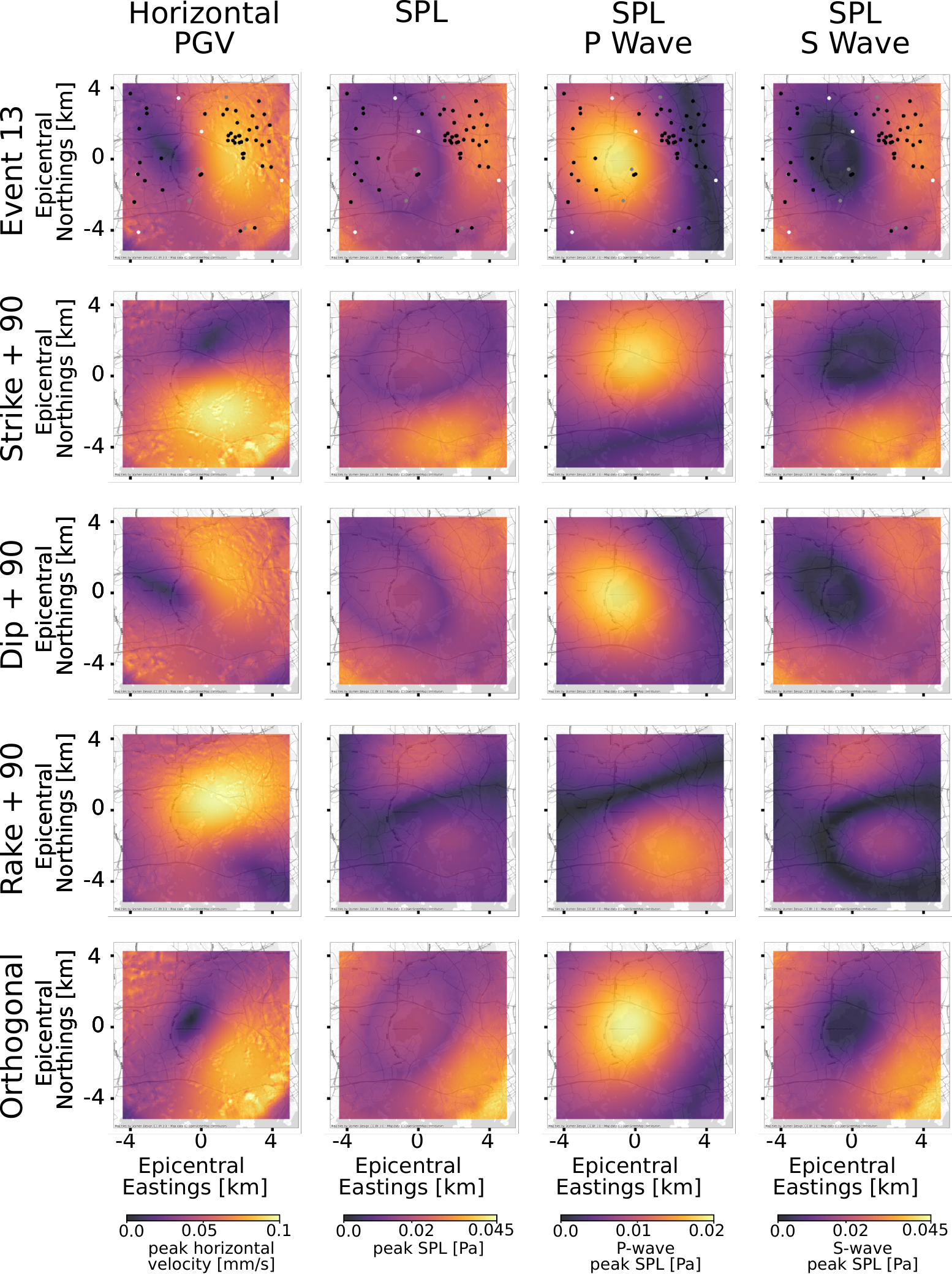}
    \caption{%
    Spatial distributions in the full model domain indicated in \cref{fig:overview}c of synthetic seismic and acoustic observables of five 3D coupled simulations using variable source geometries.
    The markers in the top row indicate observations together with the results associated with the observed \ML1.8 event mechanism. Gray, white, and black dots indicate audible, shaking, and combined sensations, respectively.
    The column \enquote{Horizontal PGV} shows the peak ground velocity in m/s.
    The column \enquote{SPL} shows the sound pressure level in Pa, estimated from peak vertical velocity and our calibration routine.
    The columns \enquote{SPL P~wave} and \enquote{SPL S~wave} show the sound pressure level in Pa estimates for the respective wave types.
    Rows two to five are associated with modified orientations of the original moment tensor point source \citep{hillers_geothermal_2020}.
    }
    \label{fig:result-maps}
\end{figure}

%
Above we presented seismo-acoustic results from five fully coupled simulations in the spatially limited high-resolution refinement area (red polygon in \cref{fig:overview}).
Now we discuss hybrid results obtained using computationally cheaper seismic-only wave propagation simulations in the larger model domain (black square in \cref{fig:overview}), with the sound pressure subsequently estimated using our scaling relation obtained from the high-resolution simulation.
For all shaking and sound distributions in \cref{fig:result-maps} we use the Webmercator projection, and we interpolate the simulated data for equidistant spacing. 
For comparison, we plot the macroseismic response distribution associated with the reference event 13 in the first row of \cref{fig:result-maps}. This is a subset of the distribution shown in \cref{fig:overview}c, where now gray, white, and black dots indicate heard, felt, and combined disturbances, respectively.
The spatial relation is discussed in \cref{sec:discussion}.
Due to our non-perfectly absorbing boundary conditions, we focus on a $\SI{8}{\km} \times \SI{8}{\km}$ area in the center of the meshed domain.
\par
The first column of \Cref{fig:result-maps} shows distributions of the peak horizontal ground velocity (PGV) using a linear \SI{}{\mm \per \s} scale for all five seismo-acoustic simulations.
Peak horizontal seismic ground velocity is a standard quantity in earthquake engineering.
In contrast to the assessment of $P$~wave and $S$~wave radiation patterns \citep{hillers_geothermal_2020}, the PGV values are a better proxy for the impinging seismic wave energy that leads to perceptible ground motions.
This is considered to fundamentally govern the public response to shaking.
The PGV maps illustrate the variable shaking intensity, and through the connection to the radiation, the PGV patterns are also controlled by the faulting mechanism.
The PGV distributions are here, to a smaller degree, modulated by topography, i.e., topographic features \enquote{light up} in our synthetic shake maps. 
This includes the north-west to south-east trending ridge at the northeastern corner of the Laajalahti bay, just south of the notorious Munkkivuori and Munkkiniemi neighborhoods, for the reference \ML1.8 event (\cref{fig:result-maps}a),
but also the hilltops in the south-east of the domain for the Strike+90 and Orthogonal scenarios.
\par
The second column shows the spatial distribution of the peak sound pressure level SPL in Pascal on a linear scale that is obtained by multiplying the spatially variable peak vertical ground velocity measured along the full synthetic wave train with the corresponding proportionality factor estimated in \cref{tab:result-linear-regression}.
These results, too, show a first order dependence on the focal mechanism, with distinct nodal zones of low or no local sound excitation at all.
We note that in these nodal areas the vertical ground velocity is zero; however, the PGV, which corresponds to the horizontal velocity, is not necessarily zero in the same regions.
\par
%
We isolate the effect of $P$~wave and $S$~wave induced sound distributions in the third and fourth column of \cref{fig:result-maps} using a linear Pascal scale for the sound pressure level.
The $P$~wave and $S$~wave patterns are complementary, regions of relatively high $P$~wave sound energy show low values of $S$~wave noise, and vice versa.
Note, however, that the choice of column-specific color ranges obscures the comparably weaker $P$~wave sounds.
The most intense earthquake sounds are associated with impinging $S$~waves.
\par
%
%
Most of the resolved source mechanisms of the induced events \citep{hillers_geothermal_2020,leonhardtSeismicityStimulation1km2021,Rintamaki2023EGU} have a very high similarity to the reference \ML1.8 event 13 mechanism in the first row in \cref{fig:result-maps}.
In the absence of weak geological zones or structures to accommodate the seismic energy release, the high similarity of the source mechanisms is governed by the fluid-induced response to the $\sim$\SI{6}{\km} deep in-situ stress conditions in the structurally homogeneous reservoir \citep{kwiatek2019controlling}.
The N\SI{110}{\degree}E direction of the maximum horizontal stress component \citep{kwiatek2019controlling} controls the dominating reverse mechanism (\cref{tab:earthquake catalogue}),
which is responsible for the predominant shaking pattern during the event sequence.
Similarly, the distribution of the SPL peak sound pressure levels in columns two to four are tied to the radiation pattern and hence to the faulting mechanism.
The distributions in rows two to five in \cref{fig:result-maps} correspond to alternative scenario focal mechanisms. 
As said, in the Helsinki case, the source mechanism variability resolved for the largest events is low \citep{Rintamaki2023EGU}, so the scenario results presumably apply only to small events, if any.
However, the variability between the scenario PGV results in the first column illustrates that a different stress regime or a different reservoir structure leading to different dominant source mechanisms significantly changes the shaking distribution.
For the audible noise estimates, the SPL peak sound pressure values in columns two to four also vary significantly between the different scenarios. This refers to the spatial distribution and to the peak values in the analyzed domain. 
The Rake+90 scenario yields similar PGV values compared to the reference event, but the excited sound shown in columns two to four is overall less powerful and hence presumably less annoying for the here considered excitation through coupling across the Earth surface, compared to the rattling of vertical structures.
The scenario results thus highlight the dependence of a geothermal stimulation soundscape on the subsurface response,
which may be taken into account by developers and regulators in addition to the more standard shaking mitigation measures.

%
%
%
\subsection{Energy}
\begin{figure}
    \centering
    \includegraphics[scale=1.0]{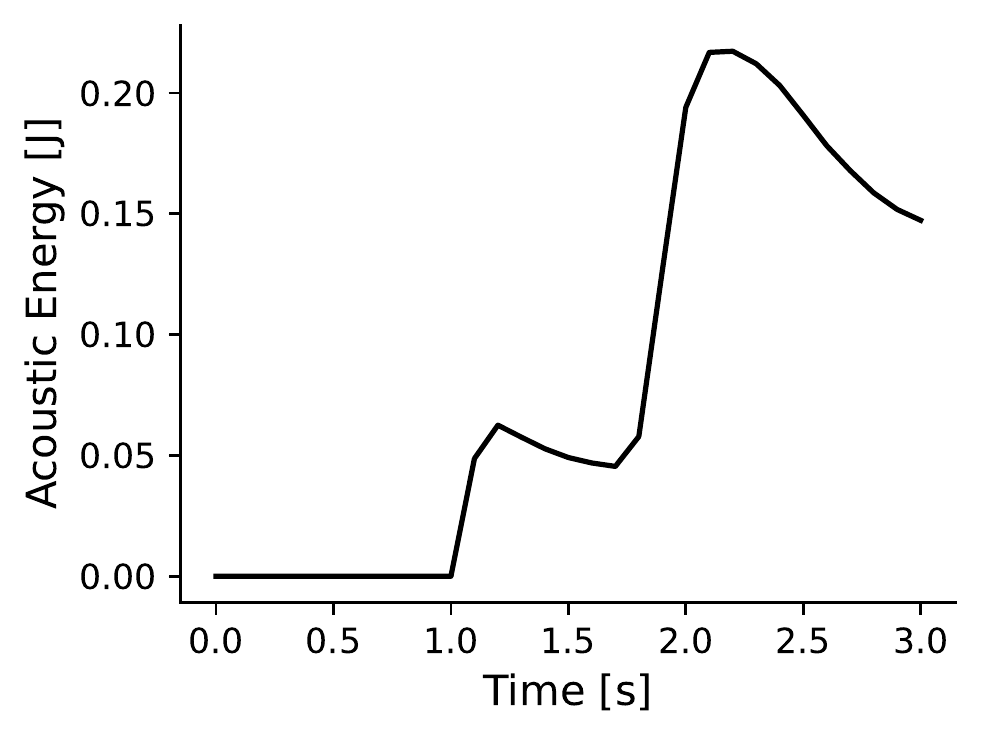}
    \caption{The acoustic energy perturbation for the reference \ML1.8 event integrated over the air layer indicates that the $S$~wave excites more acoustic energy than the $P$~wave.
    }
    \label{fig:result-energy}
\end{figure}

The dominance of the $S$~wave for sound generation can also be inferred from the evolution of the acoustic energy perturbation in our simulation.
We compute the total acoustic energy perturbation $E(t)$, which is the sum of the acoustic strain energy and the kinematic energy given by
\begin{equation}
    E(t) = \int_\Omega
    \frac{1}{2K} p^2 + 
    \frac{1}{2} \rho \Vert \bm{v} \Vert_2^2
    \, \mathrm{d}\Omega,
\end{equation}
where $\Omega$ is the total acoustic region, $K = c^2 \rho$ is the bulk modulus and $\Vert \bm{v} \Vert_2^2$ is the squared Euclidean norm of the velocity.
We only consider the perturbations in velocity and pressure and ignore the hydrostatic background pressure.
\Cref{fig:result-energy} shows the variation of the acoustic energy over time for the reference \ML1.8 event. 
Compatible with the data and synthetics in \cref{fig:fin2-comparison-receiver,fig:fin2-comparison-spectogram},
we observe that the onset of the acoustic energy perturbation coincides with the first arrival of the $P$~wave at the free surface.
The larger overall increase, however, is associated with the $S$~wave arrival.

\section{Discussion}
\label{sec:discussion}
%
%
We model the numerically and computationally challenging propagation of coupled seismic and acoustic wavefields in the Helsinki metropolitan area that were excited by stimulated small earthquakes during two EGS development stages.
The events triggered in total about 330 macroseismic responses from residents in the neighborhoods in the epicentral area on lateral scales that are similar to the $\sim$6~km depth of the stimulated seismicity.
High-quality seismic data and instrumental acoustic observations contribute to this research on the governing factors for irritable sound excitation.
Our synthetics are based on variations of the earthquake source that corresponds to the largest \ML1.8 event in 2018.
A comparison of these synthetics with seismic and acoustic data demonstrates that the employed 1D velocity model describes the subsurface structure sufficiently accurately, and that the stronger local noise excitation from impinging $S$~waves compared to $P$~waves is a robust feature of the Helsinki stimulation.
\par
%
The here presented results include frequencies up to \SI{25}{Hz}, which is at the lower limit of the human audible frequency range.
While we include realistic topography in our model, we omit the water layer in the shallow Laajalahti bay.
Including the coupling of acoustic waves across local water bodies poses a considerable meshing challenge \citep{krenz20213d}, and modeling higher frequencies is prohibitively expensive for our available computational resources.
In contrast, including low-velocity sedimentary structures that often characterize coastal regions in simulated seismic wave propagation would not noticeably increase the computational load of our approach that is controlled by the required resolution of acoustic wave propagation in air.
The upper crust of the study area in the Fennoscandian Shield is characterized by exceptionally high intrinsic $Q$ values \citep{Eulenfeld2023}, hence viscoelastic attenuation effects are negligible. Including attenuation effects in our coupled elastic-acoustic simulations is possible in the future \citep{uphoff2016generating} and may be important for the accurate modeling in sediment-rich areas.
\par
%
We evaluate the assumption that the sound pressure is given by the constant $c \rho$---sound speed times air density---times the vertical ground velocity at the Earth surface.
We fit a linear model to our synthetics from the refinement area (\cref{fig:overview}) by optimizing the prediction of peak sound pressure measured in the air layer from the peak vertical ground velocity of receivers directly below.
Overall, we achieve a good fit for our data for the main event, but
our simulated sound excitation tends to be weaker compared to the $c \rho$ scaling factor.
The regressions between modeled peak vertical velocity and peak sound pressure show that the sound excitation simulated in our fully coupled seismo-acoustic implementation can deviate from commonly employed coupling relations \citep{Tosi2000,Lamb2021}.
The quality of the parameter estimation of the linear scaling relation \cref{eq:pressure-rule-of-thumb} varies between different scenario sources.
A likely contribution to the obtained source-geometry dependence (\cref{tab:result-linear-regression}) is the variable effectiveness of the seismo-acoustic coupling that is sensitive to the interaction between radiation patterns and local topography.
It is further possible that our underestimation is related to the relatively limited horizontal domain, the steep incidence, and the $S$~wave dominance. 
\par
A consistent observation from the available single acoustic data point and from the simulated fields is that the $S$~wave excites the strongest audible sound disturbance.
This contrasts the typical assumption that the $P$~wave dominates sound generation.
Again, surface waves were not excited in the region directly above the 6~km deep reservoir.
The sample seismic data show that the $S$~wave is larger on the horizontal components, but that the vertical motion component that is more relevant for the sound excitation has also significant amplitude for the $S$~wave
(\cref{fig:result-elastic-receiver-comparison}).
The stronger $S$~wave sound excitation is also demonstrated in the acoustic data.
These observations, the synthetics, and the macroseismic reporting activity together should encourage the deployment of a sufficient number of high-frequency acoustic sensors around future induced seismicity sites.
In tandem with other sensor developments such as rotational and six-degrees-of-freedom instruments \citep{Sollberger2020}, or distributed acoustic sensing facilities \citep{Zhan2019}, more data from more diverse instrumentation can help to better understand the coupled phenomena and the environmental impact of induced earthquakes.
\par
We discuss the spatial relation of the locations of the submitted reports for the largest reference \ML1.8 event to the computed horizontal PGV shaking proxy, and to the acoustic excitation obtained from the vertical component data together with the estimated scaling factor.
For this event, 9 shaking reports were submitted, 12 on noise, and 62 reported a combination. \Cref{fig:result-maps} shows the subset in the epicentral $\SI{8}{\km} \times \SI{8}{\km}$ area.
The overall PGV distribution is ultimately governed by the radiation pattern whose $P$~wave, $SV$~wave, and $SH$~wave components are considered by \citet{hillers_geothermal_2020}.
A considerable fraction of the mixed-phenomena reports indicated by the black dots that dominate the pattern in \cref{fig:result-maps} are located in a low PGV-value region to the west of the epicenter.
Recall that these areas do sustain vertical motion---as indicated by the $P$~wave SPL sound pressure pattern in the third column---that is not captured by these horizontal PGV estimates.
The four white indicated locations of reported sound sensations and the five gray dots for shaking are insufficient to assess relations between the excitation and the reporting.
Whereas the large PGV values to the north-east suggest a likely connection between the strongest shaking in this area and the observed busiest reporting,
the variability in the synthetic PGV and SPL patterns compared to the macroseismic observations implies a more complex relation between excited wavefields and reporting activity.
\par
\citet{Tosi2000} surveyed between 500 and 2000 responses to four \Mw5.3 to \Mw5.8 events that were collected with questionnaires from employees of public authorities within 30~km around the epicenter. The Geographical Society of Finland collected 856 responses from across central Finland following a macroseismic magnitude 4.3 event on 16 November 1931 \citep{Mantyniemi2004,Mantyniemi2017}.
Close to 900 responses were collected with the Internet questionnaire after a \ML2.6 earthquake in southern Finland on 19 March 2011 \citep{Mantyniemi2017}.
This relatively high number is possibly attributed to an intermittently increased public awareness of seismic effects and readiness to share earthquake related observations, considering that the \ML2.6 event occurred eight days after the devastating \Mw9.0 Tohoku earthquake, tsunami, and Fukushima nuclear incident. Compared to these cases, the number of 83 reports associated with the largest \ML1.8 event during the 2018 Helsinki stimulation appears relatively small, which is explained by the correspondingly smaller magnitude.
In addition, the spatial distribution of the Helsinki report density varies across the affected area.
Thus, it seems premature to infer 
a single governing mechanism that explains the overall macroseismic report activity
from the visual comparison of the synthetic patterns and the felt and heard report locations.
The macroseismic observations related to the small-magnitude induced earthquakes in Otaniemi fit the overall experience from macroseismic data related to higher-magnitude non-damaging earthquakes.
\par
Macroseismic observations are affected by natural conditions such as the strength of the seismic waves, by the soil type and other site effects, by characteristics of the building type, and by the activity of the respondent at the time of observation.
The sensation can vary within a limited area, two neighbors may report quite differently.
The reporting activity can be further modulated by controversial social or other factors \citep{Mak2016,Goltz2020,Hough2021,Wald2021}, which is why the connection between our here obtained shaking and noise levels and the perceived and reported nuisance is governed by a set of very diverse variables.
The empirical connection between higher socioeconomic status and a higher tendency to report irritable phenomena in the environment in general is well documented.
Citizens in more affluent areas are typically more active,
they have more socio-cultural resources, a better understanding of governance practices, and an overall higher level of confidence that they can make a difference \citep{Arnstein2019}.
This suggests that socioeconomic factors can influence---albeit perhaps not control \citep{Wald2021}---the response patterns if we consider the reporting of transient, non-damaging environmental disturbances or nuisances as societal participation.
In summary, the observed spatial variation in the 2018 Helsinki report density can thus be governed by seismic shaking and acoustic sound effects, or, more reasonably, by a combination of these physical and social or socioeconomic effects.
Separating the physical from other effects requires the combination of survey data with more data about the physical effects obtained by data acquisition and numerical simulation such as our proposed model.
In addition, the survey data could potentially be complemented by building response types and further by a comparison of the report density and wealth or other relevant socioeconomic indicators in each neighborhood.

\section{Conclusions}
\label{sec:conc}
We presented fully coupled elastic-acoustic simulations of seismic and acoustic waves generated by 6~km deep small induced earthquakes in the Helsinki metropolitan area.
The computationally expensive simulations of seismic and sound waves across the $\SI{12}{\km} \times \SI{12}{\km}$ large epicentral area resolves frequencies around the lower limit of the human hearing sensitivity. Our setup includes realistic material models, topography, and source geometries.
We find first-order consistency between our seismic-acoustic simulations and observations. 
Our results show that shaking and sound patterns correlate with earthquake source mechanisms. 
Topographic effects can significantly influence the local acoustic wave excitation, which explains the deviations obtained from commonly applied seismo-acoustic coupling relations.
The most intense sound in the epicentral area is excited by $S$~waves.
Our comparison of the simulated seismic and acoustic wave patterns 
to the macroseismic report locations suggests that reporting activity may be linked to the peak ground velocity distribution.
The nuisance related to persistent shaking and noise associated with induced seismicity is a complex sensation that depends on the impinging wavefield properties, environment, and social and personal factors.
Together with instrumental and non-instrumental observations, multi-physics simulations can be an important tool to help understand and mitigate the mechanisms that govern public responses to shaking and noise.
The obtained relations between shaking and sound and the computational resources needed to synthesize and evaluate the scenario nuisance patterns can inform the planning of future stimulation experiments.


\paragraph{Declaration of Competing Interests}
The authors declare no competing interests.

\paragraph{Open research}
\begin{footnotesize}
We will make our computational setup available on Zenodo before publication.
\end{footnotesize}

\paragraph{Data and Resources}
\begin{footnotesize}
Data from the HE broadband monitoring stations are available through the Institute of Seismology, University of Helsinki.
Data from the 2018 temporary installations can be obtained from the GFZ German Research Centre for Geosciences Data Services \citep{HillersData2019}.
Data from the acoustic sensors are proprietary and are not released to the public.
We use topography data from the National Land Survey of Finland (\url{https://www.maanmittauslaitos.fi/}). We use the Python library \texttt{contextily} for the background maps.
The background map tiles were obtained from OpenStreet maps and Stamen Design.
The Stamen map tiles are licensed under CC BY 3.0. Data by OpenStreetMap, under ODbL.
\end{footnotesize}

\paragraph{Acknowledgments}
\begin{footnotesize}
We thank editor L.~Martire and reviewers J.~Assink and J.~W.~Bishop for comments that helped to improve the manuscript.
We thank P.~Mäntyniemi for curating the macroseismic reports, and O.~D.~Lamb and the St1 team for sharing the acoustic microphone data.
GH appreciates discussions with P.~Mäntyniemi, P.~Bäcklund, and V.~Bernelius on factors that are relevant for citizen participation. 

This work is supported by an Academy of Finland grant, decision number 337913.
The 2018 temporary deployments were supported by the Geophysical Instrument Pool Potsdam under the grants 201802.
LK, SW, AAG, and MB acknowledge funding provided (as part of the EuroHPC Joint Undertaking) for the ChEESE-2P cluster of excellence by Horizon Europe (grant agreement No.~1010930) and by the German Ministry of Research and Education. 
SW and MB acknowledge funding by the Competence Network for Scientific High Performance Computing in Bavaria (KONWIHR).
AAG acknowledges funding from the European Union’s Horizon 2020 research and innovation program (TEAR, grant agreement No.~852992), and Horizon Europe (DT-Geo, grant agreement No.~101058129 and Geo-Inquire, grant agreement No.~101058518), as well as the National Science Foundation under NSF EAR-2121666.
The CSC IT Center for Science, Finland, grand challenge project 2003841 provided access to the CSC Mahti computational infrastructure.
We acknowledge resources provided by the Gauss Centre for Supercomputing e.V.\ (\url{www.gauss-centre.eu}) on SuperMUC-NG at
the Leibniz Supercomputing Centre (\url{www.lrz.de}, project pr83no).
\end{footnotesize}

\begin{footnotesize}
\setlength{\bibsep}{1.5ex}
\bibliographystyle{tomjournal}
\bibliography{literature}

\begin{thebibliography}{91}
\providecommand{\natexlab}[1]{#1}

\bibitem[{Abercrombie(2021)}]{abercrombie2021resolution}
Abercrombie RE (2021), {Resolution and uncertainties in estimates of earthquake
  stress drop and energy release}, \emph{Philosophical Transactions of the
  Royal Society A}, 379(2196):20200131, \doi{10.1098/rsta.2020.0131}.

\bibitem[{Abrahams et~al.(2023)Abrahams, Krenz, Dunham, Gabriel, and
  Saito}]{abrahams2022comparison}
Abrahams LS, Krenz L, Dunham EM, Gabriel AA, and Saito T (2023), Comparison of
  methods for coupled earthquake and tsunami modelling, \emph{Geophysical
  Journal International}, 234(1):404--426, \doi{10.1093/gji/ggad053}.

\bibitem[{Ader et~al.(2019)Ader, Chendorain, Free, Saarno, Heikkinen, Malin,
  Leary, Kwiatek, Dresen, Bluemle, and Vuorinen}]{Ader2019}
Ader T, Chendorain M, Free M, Saarno T, Heikkinen P, Malin PE, Leary P, Kwiatek
  G, Dresen G, Bluemle F, and Vuorinen T (2019), {Design and implementation of
  a traffic light system for deep geothermal well stimulation in {Finland}},
  \emph{Journal of Seismology}, 24(5):991--1014,
  \doi{10.1007/s10950-019-09853-y}.

\bibitem[{Arnstein(2019)}]{Arnstein2019}
Arnstein SR (2019), {A Ladder of Citizen Participation}, \emph{Journal of the
  American Planning Association}, 85(1):24--34,
  \doi{10.1080/01944363.2018.1559388}.

\bibitem[{Arrowsmith et~al.(2010)Arrowsmith, Johnson, Drob, and
  Hedlin}]{Arrowsmith2010}
Arrowsmith SJ, Johnson JB, Drob DP, and Hedlin MAH (2010), {The seismoacoustic
  wavefield: A new paradigm in studying geophysical phenomena}, \emph{Reviews
  of Geophysics}, 48(4), \doi{10.1029/2010RG000335}.

\bibitem[{Averbuch et~al.(2020)Averbuch, Assink, and
  Evers}]{averbuchLongrangeAtmosphericInfrasound2020}
Averbuch G, Assink JD, and Evers LG (2020), Long-range atmospheric infrasound
  propagation from subsurface sources, \emph{The Journal of the Acoustical
  Society of America}, 147(2):1264--1274, \doi{10.1121/10.0000792}.

\bibitem[{Baisch et~al.(2019)Baisch, Koch, and Muntendam‐Bos}]{Baisch2019}
Baisch S, Koch C, and Muntendam‐Bos A (2019), Traffic light systems: To what
  extent can induced seismicity be controlled?, \emph{Seismological Research
  Letters}, 90(3):1145--1154, \doi{10.1785/0220180337}.

\bibitem[{Bentz et~al.(2020)Bentz, Kwiatek, Martínez-Garzón, Bohnhoff, and
  Dresen}]{Bentz2020}
Bentz S, Kwiatek G, Martínez-Garzón P, Bohnhoff M, and Dresen G (2020),
  Seismic moment evolution during hydraulic stimulations, \emph{Geophysical
  Research Letters}, 47(5):e2019GL086185, \doi{10.1029/2019GL086185},
  e2019GL086185 2019GL086185.

\bibitem[{Bommer et~al.(2006)Bommer, Oates, Cepeda, Lindholm, Bird, Torres,
  Marroquín, and Rivas}]{Bommer2006}
Bommer JJ, Oates S, Cepeda JM, Lindholm C, Bird J, Torres R, Marroquín G, and
  Rivas J (2006), {Control of hazard due to seismicity induced by a hot
  fractured rock geothermal project}, \emph{Engineering Geology},
  83(4):287--306, \doi{10.1016/j.enggeo.2005.11.002}.

\bibitem[{Bommer et~al.(2017)Bommer, Stafford, Edwards, Dost, {van Dedem},
  {Rodriguez-Marek}, Kruiver, {van Elk}, Doornhof, and
  Ntinalexis}]{bommerFrameworkGroundMotionModel2017}
Bommer JJ, Stafford PJ, Edwards B, Dost B, {van Dedem} E, {Rodriguez-Marek} A,
  Kruiver P, {van Elk} J, Doornhof D, and Ntinalexis M (2017), Framework for a
  ground-motion model for induced seismic hazard and risk analysis in the
  {Groningen} gas field, {{The Netherlands}}, \emph{Earthquake Spectra},
  33(2):481--498, \doi{10.1193/082916EQS138M}.

\bibitem[{Breuer et~al.(2016)Breuer, Heinecke, and
  Bader}]{breuer_petascale_2016}
Breuer A, Heinecke A, and Bader M (2016), Petascale local time stepping for the
  {ADER}-{DG} finite element method, in 2016 {IEEE} {International} {Parallel}
  and {Distributed} {Processing} {Symposium} ({IPDPS}), pages 854--863,
  \doi{10.1109/IPDPS.2016.109}.

\bibitem[{Brissaud et~al.(2021)Brissaud, Krishnamoorthy, Jackson, Bowman,
  Komjathy, Cutts, Zhan, Pauken, Izraelevitz, and Walsh}]{Brissaud2021}
Brissaud Q, Krishnamoorthy S, Jackson JM, Bowman DC, Komjathy A, Cutts JA, Zhan
  Z, Pauken MT, Izraelevitz JS, and Walsh GJ (2021), The first detection of an
  earthquake from a balloon using its acoustic signature, \emph{Geophysical
  Research Letters}, 48(12), \doi{10.1029/2021gl093013}.

\bibitem[{Brissaud et~al.(2017)Brissaud, Martin, Garcia, and
  Komatitsch}]{Brissaud2017}
Brissaud Q, Martin R, Garcia RF, and Komatitsch D (2017), Hybrid {Galerkin}
  numerical modelling of elastodynamics and compressible {Navier}-{Stokes}
  couplings: applications to seismo-gravito acoustic waves, \emph{Geophysical
  Journal International}, 210(2):1047--1069, \doi{10.1093/gji/ggx185}.

\bibitem[{Brooks et~al.(2018)Brooks, Stein, Spencer, Salditch, Petersen, and
  McNamara}]{Brooks2018}
Brooks EM, Stein S, Spencer BD, Salditch L, Petersen MD, and McNamara DE
  (2018), {Assessing earthquake hazard map performance for natural and induced
  seismicity in the central and eastern United States}, \emph{Seismological
  Research Letters}, 89(1):118--126.

\bibitem[{Brune(1970)}]{bruneTectonicStressSpectra1970}
Brune JN (1970), Tectonic stress and the spectra of seismic shear waves from
  earthquakes, \emph{Journal of Geophysical Research (1896-1977)},
  75(26):4997--5009, \doi{10.1029/JB075i026p04997}.

\bibitem[{Che et~al.(2022)Che, Kim, Le~Pichon, Park, Arrowsmith, and
  Stump}]{cheIlluminatingNorthKorean2022}
Che IY, Kim K, Le~Pichon A, Park J, Arrowsmith S, and Stump B (2022),
  Illuminating the {{North Korean}} nuclear explosion test in 2017 using remote
  infrasound observations, \emph{Geophysical Journal International},
  228(1):308--315, \doi{10.1093/gji/ggab338}.

\bibitem[{Cook(1971)}]{cookInfrasoundRadiatedMontana1971}
Cook RK (1971), Infrasound radiated during the {Montana} earthquake of 1959
  {August} 18, \emph{Geophysical Journal International}, 26(1-4):191--198,
  \doi{10.1111/j.1365-246X.1971.tb03393.x}.

\bibitem[{Courant et~al.(1928)Courant, Friedrichs, and
  Lewy}]{courantUeberPartiellenDifferenzengleichungen1928}
Courant R, Friedrichs K, and Lewy H (1928), {\"Uber die partiellen
  Differenzengleichungen der mathematischen Physik}, \emph{Mathematische
  Annalen}, 100(1):32--74, \doi{10.1007/BF01448839}.

\bibitem[{Davison(1938)}]{Davison1938}
Davison C (1938), {Earthquake sounds}, \emph{Bulletin of the Seismological
  Society of America}, 28(3):147--161, \doi{10.1785/BSSA0280030147}.

\bibitem[{Dumbser and
  K{\"a}ser(2006)}]{dumbserArbitraryHighorderDiscontinuous2006}
Dumbser M and K{\"a}ser M (2006), An arbitrary high-order discontinuous
  {{Galerkin}} method for elastic waves on unstructured meshes - {{II}}.
  {{The}} three-dimensional isotropic case, \emph{Geophysical Journal
  International}, 167(1):319--336, \doi{10.1111/j.1365-246X.2006.03120.x}.

\bibitem[{Ebel et~al.(1982)Ebel, Vudler, and Celata}]{Ebel1982}
Ebel JE, Vudler V, and Celata M (1982), {The 1981 microearthquake swarm near
  {Moodus}, {Connecticut}}, \emph{Geophysical Research Letters}, 9(4):397--400,
  \doi{10.1029/GL009i004p00397}.

\bibitem[{Edwards et~al.(2014)Edwards, de~Groot‐Hedlin, and
  Hedlin}]{Edwards2014}
Edwards WN, de~Groot‐Hedlin CD, and Hedlin MAH (2014), Forensic investigation
  of a probable meteor sighting using {USArray} acoustic data,
  \emph{Seismological Research Letters}, 85(5):1012--1018,
  \doi{10.1785/0220140056}.

\bibitem[{Eulenfeld et~al.(2023)Eulenfeld, Hillers, Vuorinen, and
  Wegler}]{Eulenfeld2023}
Eulenfeld T, Hillers G, Vuorinen TAT, and Wegler U (2023), {Induced Earthquake
  Source Parameters, Attenuation, and Site Effects From Waveform Envelopes in
  the Fennoscandian Shield}, \emph{Journal of Geophysical Research: Solid
  Earth}, 128(4):e2022JB025162, \doi{https://doi.org/10.1029/2022JB025162},
  e2022JB025162 2022JB025162.

\bibitem[{Evers et~al.(2014)Evers, Brown, Heaney, Assink, Smets, and
  Snellen}]{Evers2014}
Evers LG, Brown D, Heaney KD, Assink JD, Smets PSM, and Snellen M (2014),
  {Evanescent wave coupling in a geophysical system: Airborne acoustic signals
  from the Mw 8.1 Macquarie Ridge earthquake}, \emph{Geophysical Research
  Letters}, 41(5):1644--1650, \doi{10.1002/2013GL058801}.

\bibitem[{Fastl and Zwicker(2006)}]{fastl2006psychoacoustics}
Fastl H and Zwicker E (2006), Psychoacoustics: facts and models, volume~22,
  Springer Science \& Business Media, \doi{10.1007/978-3-540-68888-4}.

\bibitem[{Galis et~al.(2017)Galis, Ampuero, Mai, and
  Cappa}]{galisInducedSeismicityProvides2017}
Galis M, Ampuero JP, Mai PM, and Cappa F (2017), Induced seismicity provides
  insight into why earthquake ruptures stop, \emph{Science Advances},
  3(12):eaap7528, \doi{10.1126/sciadv.aap7528}.

\bibitem[{Gaucher et~al.(2015)Gaucher, Schoenball, Heidbach, Zang, Fokker, {van
  Wees}, and Kohl}]{Gaucher2015}
Gaucher E, Schoenball M, Heidbach O, Zang A, Fokker PA, {van Wees} JD, and Kohl
  T (2015), {Induced seismicity in geothermal reservoirs: A review of
  forecasting approaches}, \emph{Renewable and Sustainable Energy Reviews},
  52:1473--1490, \doi{10.1016/j.rser.2015.08.026}.

\bibitem[{Goltz et~al.(2020)Goltz, Park, Nakano, and Yamori}]{Goltz2020}
Goltz JD, Park H, Nakano G, and Yamori K (2020), {Earthquake ground motion and
  human behavior: Using DYFI data to assess behavioral response to
  earthquakes}, \emph{Earthquake Spectra}, 36(3):1231--1253.

\bibitem[{Grigoli et~al.(2017)Grigoli, Cesca, Priolo, Rinaldi, Clinton,
  Stabile, Dost, Fernandez, Wiemer, and Dahm}]{Grigoli2017}
Grigoli F, Cesca S, Priolo E, Rinaldi AP, Clinton JF, Stabile TA, Dost B,
  Fernandez MG, Wiemer S, and Dahm T (2017), {Current challenges in monitoring,
  discrimination, and management of induced seismicity related to underground
  industrial activities: A European perspective}, \emph{Reviews of Geophysics},
  55(2):310--340, \doi{10.1002/2016RG000542}.

\bibitem[{H{\"a}ring et~al.(2008)H{\"a}ring, Schanz, Ladner, and
  Dyer}]{Haring2008}
H{\"a}ring MO, Schanz U, Ladner F, and Dyer BC (2008), {C}haracterisation of
  the {B}asel 1 enhanced geothermal system, \emph{Geothermics}, 37:469--495,
  \doi{10.1016/j.geothermics.2008.06.002}.

\bibitem[{Hedlin et~al.(2012)Hedlin, Walker, Drob, and
  de~Groot-Hedlin}]{hedlin2012infrasound}
Hedlin M, Walker K, Drob D, and de~Groot-Hedlin C (2012), {Infrasound:
  Connecting the solid earth, oceans, and atmosphere}, \emph{Annual Review of
  Earth and Planetary Sciences}, 40(327):2012,
  \doi{10.1146/annurev-earth-042711-105508}.

\bibitem[{Hernandez et~al.(2018)Hernandez, Le~Pichon, Vergoz, Herry, Ceranna,
  Pilger, Marchetti, Ripepe, and Bossu}]{hernandezEstimatingGroundMotion2018}
Hernandez B, Le~Pichon A, Vergoz J, Herry P, Ceranna L, Pilger C, Marchetti E,
  Ripepe M, and Bossu R (2018), Estimating the ground-motion distribution of
  the 2016 {{Mw}} 6.2 {{Amatrice}}, {{Italy}}, earthquake using remote
  infrasound observations, \emph{Seismological Research Letters},
  89(6):2227--2236, \doi{10.1785/0220180103}.

\bibitem[{Hesthaven and Warburton(2008)}]{hesthaven_nodal_2008}
Hesthaven JS and Warburton T (2008), Nodal {Discontinuous} {Galerkin}
  {Methods}: {Algorithms}, {Analysis}, and {Applications}, Texts in {Applied}
  {Mathematics}, Springer-Verlag, New York, \doi{10.1007/978-0-387-72067-8}.

\bibitem[{Hill(2011)}]{hill2011}
Hill DP (2011), {What is that mysterious booming sound?}, \emph{Seismological
  Research Letters}, 82(5):619--622, \doi{doi:10.1785/gssrl.82.5.619}.

\bibitem[{Hill et~al.(1976)Hill, Fischer, Lahr, and
  Coakley}]{hill1976earthquake}
Hill DP, Fischer FG, Lahr KM, and Coakley JM (1976), {Earthquake sounds
  generated by body-wave ground motion}, \emph{Bulletin of the Seismological
  Society of America}, 66(4):1159--1172, \doi{10.1785/BSSA0660041159}.

\bibitem[{Hillers et~al.(2019)Hillers, Vuorinen, Arola, Katajisto, Koskenniemi,
  McKevitt, Rezaei, Rinne, Salmenperä, Seipäjärvi, Väkevä, Voutilainen,
  Arhe, Juntunen, Keskinen, Lindblom, Oinonen, and Tiira}]{HillersData2019}
Hillers G, Vuorinen TAT, Arola EJ, Katajisto VE, Koskenniemi MP, McKevitt BM,
  Rezaei S, Rinne LA, Salmenperä IE, Seipäjärvi PJ, Väkevä LSO,
  Voutilainen AI, Arhe K, Juntunen AK, Keskinen J, Lindblom PY, Oinonen K, and
  Tiira T (2019), {A 100 3-component sensor deployment to monitor the 2018
  {EGS} stimulation in {Espoo}/{Helsinki}, southern {Finland} - {Datasets}},
  \doi{10.5880/GIPP.201802.1}.

\bibitem[{Hillers et~al.(2020)Hillers, Vuorinen, Uski, Kortström, Mäntyniemi,
  Tiira, Malin, and Saarno}]{hillers_geothermal_2020}
Hillers G, Vuorinen TAT, Uski MR, Kortström JT, Mäntyniemi PB, Tiira T, Malin
  PE, and Saarno T (2020), The 2018 geothermal reservoir stimulation in
  {Espoo}/{Helsinki}, southern {Finland}: Seismic network anatomy and data
  features, \emph{Seismological Research Letters}, \doi{10.1785/0220190253}.

\bibitem[{Holmgren et~al.(2023)Holmgren, Kwiatek, and Werner}]{Holmgren2023}
Holmgren JM, Kwiatek G, and Werner MJ (2023), Nonsystematic rupture directivity
  of geothermal energy induced microseismicity in helsinki, finland,
  \emph{Journal of Geophysical Research: Solid Earth}, 128(3):e2022JB025226,
  \doi{https://doi.org/10.1029/2022JB025226}, e2022JB025226 2022JB025226.

\bibitem[{Hough and Martin(2021)}]{Hough2021}
Hough SE and Martin SS (2021), {Which Earthquake Accounts Matter?},
  \emph{Seismological Research Letters}, 92(2A):1069--1084,
  \doi{10.1785/0220200366}.

\bibitem[{Hupe et~al.(2022)Hupe, Ceranna, Le~Pichon, Matoza, and
  Mialle}]{Hupe2022}
Hupe P, Ceranna L, Le~Pichon A, Matoza RS, and Mialle P (2022), {International
  Monitoring System infrasound data products for atmospheric studies and
  civilian applications}, \emph{Earth System Science Data}, 14(9):4201--4230,
  \doi{10.5194/essd-14-4201-2022}.

\bibitem[{Keil et~al.(2022)Keil, Wassermann, and Megies}]{Keil2022}
Keil S, Wassermann J, and Megies T (2022), {Estimation of ground motion due to
  induced seismicity at a geothermal power plant near Munich, Germany, using
  numerical simulations}, \emph{Geothermics}, 106:102577.

\bibitem[{Krenz et~al.(2021)Krenz, Uphoff, Ulrich, Gabriel, Abrahams, Dunham,
  and Bader}]{krenz20213d}
Krenz L, Uphoff C, Ulrich T, Gabriel AA, Abrahams LS, Dunham EM, and Bader M
  (2021), {3D acoustic-elastic coupling with gravity: the dynamics of the 2018
  {Palu}, {Sulawesi} earthquake and tsunami}, in Proceedings of the
  International Conference for High Performance Computing, Networking, Storage
  and Analysis, pages 1--14, \doi{10.1145/3458817.3476173}.

\bibitem[{Kwiatek et~al.(2022)Kwiatek, Martinez~Garzon, Davidsen, Malin,
  Karjalainen, Bohnhoff, and Dresen}]{Kwiatek2022}
Kwiatek G, Martinez~Garzon P, Davidsen J, Malin P, Karjalainen A, Bohnhoff M,
  and Dresen G (2022), {Limited earthquake interaction during a geothermal
  hydraulic stimulation in Helsinki, Finland}, \emph{Journal of Geophysical
  Research: Solid Earth}, 127(9), \doi{10.1029/2022JB024354}.

\bibitem[{Kwiatek et~al.(2019)Kwiatek, Saarno, Ader, Bluemle, Bohnhoff,
  Chendorain, Dresen, Heikkinen, Kukkonen, Leary, Leonhardt, Malin,
  Mart{\'{\i}}nez-Garz{\'{o}}n, Passmore, Passmore, Valenzuela, and
  Wollin}]{kwiatek2019controlling}
Kwiatek G, Saarno T, Ader T, Bluemle F, Bohnhoff M, Chendorain M, Dresen G,
  Heikkinen P, Kukkonen I, Leary P, Leonhardt M, Malin P,
  Mart{\'{\i}}nez-Garz{\'{o}}n P, Passmore K, Passmore P, Valenzuela S, and
  Wollin C (2019), {Controlling fluid-induced seismicity during a 6.1-km-deep
  geothermal stimulation in Finland}, \emph{Science Advances}, 5(5):eaav7224,
  \doi{10.1126/sciadv.aav7224}.

\bibitem[{Käser et~al.(2008)Käser, Hermann, and
  Puente}]{kaser_quantitative_2008}
Käser M, Hermann V, and Puente Jdl (2008), Quantitative accuracy analysis of
  the discontinuous {Galerkin} method for seismic wave propagation,
  \emph{Geophysical Journal International}, 173(3):990--999,
  \doi{10.1111/j.1365-246X.2008.03781.x}.

\bibitem[{Lamb et~al.(2021)Lamb, Lees, Malin, and Saarno}]{Lamb2021}
Lamb OD, Lees JM, Malin PE, and Saarno T (2021), {Audible acoustics from
  low-magnitude fluid-induced earthquakes in Finland}, \emph{Scientific
  reports}, 11(1):1--8, \doi{10.1038/s41598-021-98701-6}.

\bibitem[{Lecoulant et~al.(2019)Lecoulant, Guennou, Guillon, and
  Royer}]{Lecoulant2019}
Lecoulant J, Guennou C, Guillon L, and Royer JY (2019), {Three-dimensional
  modeling of earthquake generated acoustic waves in the ocean in simplified
  configurations}, \emph{The Journal of the Acoustical Society of America},
  146(3):2113--2123, \doi{10.1121/1.5126009}.

\bibitem[{Leonhardt et~al.(2021)Leonhardt, Kwiatek, {Mart{\'i}nez-Garz{\'o}n},
  Bohnhoff, Saarno, Heikkinen, and
  Dresen}]{leonhardtSeismicityStimulation1km2021}
Leonhardt M, Kwiatek G, {Mart{\'i}nez-Garz{\'o}n} P, Bohnhoff M, Saarno T,
  Heikkinen P, and Dresen G (2021), Seismicity during and after stimulation of
  a 6.1km deep enhanced geothermal system in {{Helsinki}}, {{Finland}},
  \emph{Solid Earth}, 12(3):581--594, \doi{10.5194/se-12-581-2021}.

\bibitem[{Li et~al.(2021)Li, Gabriel, Rintam{\"a}ki, and Hillers}]{Li2021EGU}
Li B, Gabriel AA, Rintam{\"a}ki A, and Hillers G (2021), {Array based analysis
  of induced earthquake characteristics using beamforming and back-projection
  methods in Helsinki, Finland}, in EGU General Assembly Conference Abstracts,
  pages EGU21--12888, \doi{10.5194/egusphere-egu21-12888}.

\bibitem[{Li et~al.(2022)Li, Wu, Bao, Oglesby, Ghosh, Gabriel, Meng, and
  Chu}]{li2022rupture}
Li B, Wu B, Bao H, Oglesby DD, Ghosh A, Gabriel AA, Meng L, and Chu R (2022),
  Rupture heterogeneity and directivity effects in back-projection analysis,
  \emph{Journal of Geophysical Research: Solid Earth}, 127(3):e2021JB022663,
  \doi{10.1029/2021JB022663}.

\bibitem[{Lund et~al.(2015)Lund, Malm, M{\"a}ntyniemi, Oinonen, Tiira, Uski,
  and Vuorinen}]{saari2015evaluating}
Lund B, Malm M, M{\"a}ntyniemi P, Oinonen K, Tiira T, Uski M, and Vuorinen T
  (2015), Evaluating seismic hazard for the {Hanhikivi} nuclear power plant
  site, seismological characteristics of the source areas, attenuation of
  seismic signal, and probabilistic analysis of seismic hazard., volume Report
  NE-4459, {\AA}F-Consult Ltd, Finland.

\bibitem[{Madariaga(2011)}]{madariagaEarthquakeScalingLaws2011}
Madariaga R (2011), Earthquake scaling laws, in RA~Meyers, editor, Extreme
  {{Environmental Events}}, pages 364--383, {Springer New York}, {New York,
  NY}, \doi{10.1007/978-1-4419-7695-6_22}.

\bibitem[{Majer et~al.(2012)Majer, Nelson, Robertson-Tait, Savy, and
  Wong}]{majer2012protocol}
Majer E, Nelson J, Robertson-Tait A, Savy J, and Wong I (2012), {Protocol for
  addressing induced seismicity associated with enhanced geothermal systems},
  \emph{US Department of Energy}, page~52.

\bibitem[{Mak and Schorlemmer(2016)}]{Mak2016}
Mak S and Schorlemmer D (2016), {What Makes People Respond to “Did You Feel
  It?”?}, \emph{Seismological Research Letters}, 87(1):119--131,
  \doi{10.1785/0220150056}.

\bibitem[{M{\"a}ntyniemi(2004)}]{Mantyniemi2004}
M{\"a}ntyniemi PB (2004), {Pre-instrumental earthquakes in a low-seismicity
  region: A reinvestigation of the macroseismic data for the 16 November 1931
  events in Central Finland using statistical analysis}, \emph{Journal of
  seismology}, 8(1):71--90, \doi{10.1023/B:JOSE.0000009501.13091.2d}.

\bibitem[{M{\"a}ntyniemi(2017)}]{Mantyniemi2017}
M{\"a}ntyniemi PB (2017), {Macroseismology in Finland from the 1730s to the
  2000s. Part 2: From an obligation of the learned elite to citizen science},
  \emph{Geophysica}, 52:23--41.

\bibitem[{M{\"a}ntyniemi(2022)}]{Mantyniemi2022}
M{\"a}ntyniemi PB (2022), Revisiting {Svenskby}, southeastern {Finland}:
  Communications regarding low-magnitude earthquakes in 1751--1752,
  \emph{Geosciences}, 12(9), \doi{10.3390/geosciences12090338}.

\bibitem[{Megies and Wassermann(2014)}]{Megies2014}
Megies T and Wassermann J (2014), {Microseismicity observed at a
  non-pressure-stimulated geothermal power plant}, \emph{Geothermics},
  52:36--49, \doi{10.1016/j.geothermics.2014.01.002}, analysis of Induced
  Seismicity in Geothermal Operations.

\bibitem[{Mignan et~al.(2017)Mignan, Broccardo, Wiemer, and
  Giardini}]{Mignan2017}
Mignan A, Broccardo M, Wiemer S, and Giardini D (2017), {Induced seismicity
  closed-form traffic light system for actuarial decision-making during deep
  fluid injections}, \emph{Scientific reports}, 7(1):1--10,
  \doi{10.1038/s41598-017-13585-9}.

\bibitem[{M{\o}ller and Pedersen(2004)}]{moller2004hearing}
M{\o}ller H and Pedersen CS (2004), {Hearing at low and infrasonic
  frequencies}, \emph{Noise \& Health}, 6(23):37--57.

\bibitem[{Mutschlecner and Whitaker(2005)}]{Mutschlecner2005}
Mutschlecner JP and Whitaker RW (2005), {Infrasound from earthquakes},
  \emph{Journal of Geophysical Research: Atmospheres}, 110(D1):1--11,
  \doi{10.1029/2004JD005067}.

\bibitem[{Mühlhans(2017)}]{muhlhans2017infrasound}
Mühlhans JH (2017), {Low frequency and infrasound: A critical review of the
  myths, misbeliefs and their relevance to music perception research},
  \emph{Musicae Scientiae}, 21(3):267--286, \doi{10.1177/1029864917690931}.

\bibitem[{Pitarka et~al.(2022)Pitarka, Akinci, De~Gori, and
  Buttinelli}]{Pitarka2022}
Pitarka A, Akinci A, De~Gori P, and Buttinelli M (2022), Deterministic {3D}
  ground-motion simulations (0--5~{Hz}) and surface topography effects of the
  30 {October} 2016 {Mw}6.5 {Norcia}, {Italy}, earthquake, \emph{Bulletin of
  the Seismological Society of America}, 112(1):262--286,
  \doi{10.1785/0120210133}.

\bibitem[{Rintamäki et~al.(2023)Rintamäki, Hillers, Heimann, Dahm, and
  Korja}]{Rintamaki2023EGU}
Rintamäki A, Hillers G, Heimann S, Dahm T, and Korja A (2023), {Centroid full
  moment tensor analysis reveals fluid channels opened by induced seismicity at
  EGS, Helsinki region, southern Finland}, in EGU General Assembly Conference
  Abstracts, pages EGU23--12756, \doi{10.5194/egusphere-egu23-12756}.

\bibitem[{Rintamäki et~al.(2021)Rintamäki, Hillers, Vuorinen, Luhta, Pownall,
  Tsarsitalidou, Galvin, Keskinen, Kortström, Lin, Mäntyniemi, Oinonen,
  Oksanen, Seipäjärvi, Taylor, Uski, Voutilainen, and Whipp}]{Rintamaki2021}
Rintamäki AE, Hillers G, Vuorinen TAT, Luhta T, Pownall JM, Tsarsitalidou C,
  Galvin K, Keskinen J, Kortström JT, Lin TC, Mäntyniemi PB, Oinonen KJ,
  Oksanen TJ, Seipäjärvi PJ, Taylor G, Uski MR, Voutilainen AI, and Whipp DM
  (2021), {A seismic network to monitor the 2020 {EGS} stimulation in the
  {Espoo}/{Helsinki} area, southern {Finland}}, \emph{Seismological Research
  Letters}, 93(2A):1046--1062, \doi{10.1785/0220210195}.

\bibitem[{Rutqvist et~al.(2014)Rutqvist, Cappa, Rinaldi, and
  Godano}]{Rutqvist2014perception}
Rutqvist J, Cappa F, Rinaldi AP, and Godano M (2014), {Modeling of induced
  seismicity and ground vibrations associated with geologic CO2 storage, and
  assessing their effects on surface structures and human perception},
  \emph{International Journal of Greenhouse Gas Control}, 24:64--77,
  \doi{10.1016/j.ijggc.2014.02.017}.

\bibitem[{Schultz et~al.(2021{\natexlab{a}})Schultz, Beroza, and
  Ellsworth}]{Schultz2021a}
Schultz R, Beroza GC, and Ellsworth WL (2021{\natexlab{a}}), {A risk-based
  approach for managing hydraulic fracturing--induced seismicity},
  \emph{Science}, 372(6541):504--507, \doi{10.1126/science.abg5451}.

\bibitem[{Schultz et~al.(2021{\natexlab{b}})Schultz, Quitoriano, Wald, and
  Beroza}]{Schultz2021b}
Schultz R, Quitoriano V, Wald DJ, and Beroza GC (2021{\natexlab{b}}),
  {Quantifying nuisance ground motion thresholds for induced earthquakes},
  \emph{Earthquake Spectra}, 37(2):789--802, \doi{10.1177/8755293020988025}.

\bibitem[{Seabold and Perktold(2010)}]{seabold2010statsmodels}
Seabold S and Perktold J (2010), Statsmodels: Econometric and statistical
  modeling with {Python}, in 9th Python in Science Conference,
  \doi{10.25080/Majora-92bf1922-011}.

\bibitem[{{Shani-Kadmiel} et~al.(2021){Shani-Kadmiel}, Averbuch, Smets, Assink,
  and Evers}]{shani-kadmiel2010HaitiEarthquake2021}
{Shani-Kadmiel} S, Averbuch G, Smets P, Assink J, and Evers L (2021), The 2010
  {{Haiti}} earthquake revisited: {{An}} acoustic intensity map from remote
  atmospheric infrasound observations, \emph{Earth and Planetary Science
  Letters}, 560:116795, \doi{10.1016/j.epsl.2021.116795}.

\bibitem[{Smink et~al.(2019)Smink, Assink, Bosveld, Smets, and
  Evers}]{sminkThreeDimensionalArrayStudy2019}
Smink MME, Assink JD, Bosveld FC, Smets PSM, and Evers LG (2019), A
  three-dimensional array for the study of infrasound propagation through the
  atmospheric boundary layer, \emph{Journal of Geophysical Research:
  Atmospheres}, 124(16):9299--9313, \doi{10.1029/2019JD030386}.

\bibitem[{Sollberger et~al.(2020)Sollberger, Igel, Schmelzbach, Edme, van
  Manen, Bernauer, Yuan, Wassermann, Schreiber, and
  Robertsson}]{Sollberger2020}
Sollberger D, Igel H, Schmelzbach C, Edme P, van Manen DJ, Bernauer F, Yuan S,
  Wassermann J, Schreiber U, and Robertsson JOA (2020), {Seismological
  Processing of Six Degree-of-Freedom Ground-Motion Data}, \emph{Sensors},
  20(23), \doi{10.3390/s20236904}.

\bibitem[{Stauffacher et~al.(2015)Stauffacher, Muggli, Scolobig, and
  Moser}]{Stauffacher2015}
Stauffacher M, Muggli N, Scolobig A, and Moser C (2015), Framing deep
  geothermal energy in mass media: The case of {Switzerland},
  \emph{Technological Forecasting and Social Change}, 98:60--70,
  \doi{10.1016/j.techfore.2015.05.018}.

\bibitem[{Sylvander and Mogos(2005)}]{Sylvander2005}
Sylvander M and Mogos DG (2005), The sounds of small earthquakes: Quantitative
  results from a study of regional macroseismic bulletins, \emph{Bulletin of
  the Seismological Society of America}, 95(4):1510--1515,
  \doi{10.1785/0120040197}.

\bibitem[{Sylvander et~al.(2007)Sylvander, Ponsolles, Benahmed, and
  Fels}]{Sylvander2007}
Sylvander M, Ponsolles C, Benahmed S, and Fels JF (2007), Seismoacoustic
  recordings of small earthquakes in the {Pyrenees}: Experimental results,
  \emph{Bulletin of the Seismological Society of America}, 97(1B):294--304,
  \doi{10.1785/0120060009}.

\bibitem[{Taufiqurrahman et~al.(2022)Taufiqurrahman, Gabriel, Ulrich,
  Valentova, and Gallovič}]{Taufiqurrahman2022}
Taufiqurrahman T, Gabriel AA, Ulrich T, Valentova L, and Gallovič F (2022),
  {Broadband dynamic rupture modeling with fractal fault roughness, frictional
  heterogeneity, viscoelasticity and topography: the 2016 Mw 6.2 Amatrice,
  Italy earthquake}, \emph{Geophysical Research Letters}, 49(22),
  \doi{https://doi.org/10.1029/2022GL098872}.

\bibitem[{Taylor et~al.(2021)Taylor, Hillers, and Vuorinen}]{Taylor2021}
Taylor G, Hillers G, and Vuorinen TAT (2021), Using array-derived rotational
  motion to obtain local wave propagation properties from earthquakes induced
  by the 2018 geothermal stimulation in {Finland}, \emph{Geophysical Research
  Letters}, 48(6):e2020GL090403, \doi{10.1029/2020GL090403}, e2020GL090403
  2020GL090403.

\bibitem[{Thouvenot et~al.(2009)Thouvenot, Jenatton, and
  Gratier}]{Thouvenot2009}
Thouvenot F, Jenatton L, and Gratier JP (2009), {200-m-deep earthquake swarm in
  Tricastin (lower Rhône Valley, France) accounts for noisy seismicity over
  past centuries}, \emph{Terra Nova}, 21(3):203--210,
  \doi{10.1111/j.1365-3121.2009.00875.x}.

\bibitem[{Tomic et~al.(2009)Tomic, Abercrombie, and
  Do~Nascimento}]{tomic2009source}
Tomic J, Abercrombie R, and Do~Nascimento A (2009), {Source parameters and
  rupture velocity of small M$\leq$2.1 reservoir induced earthquakes},
  \emph{Geophysical Journal International}, 179(2):1013--1023,
  \doi{10.1111/j.1365-246X.2009.04233.x}.

\bibitem[{Tosi et~al.(2000)Tosi, De~Rubeis, Tertulliani, and
  Gasparini}]{Tosi2000}
Tosi P, De~Rubeis V, Tertulliani A, and Gasparini C (2000), {Spatial patterns
  of earthquake sounds and seismic source geometry}, \emph{Geophysical Research
  Letters}, 27(17):2749--2752, \doi{10.1029/2000GL011377}.

\bibitem[{Ulrich et~al.(2022)Ulrich, Gabriel, and Madden}]{Ulrich2022}
Ulrich T, Gabriel AA, and Madden EH (2022), Stress, rigidity and sediment
  strength control megathrust earthquake and tsunami dynamics, \emph{Nature
  Geoscience}, 15(1):67--73, \doi{10.1038/s41561-021-00863-5}.

\bibitem[{Uphoff and Bader(2016)}]{uphoff2016generating}
Uphoff C and Bader M (2016), {Generating high performance matrix kernels for
  earthquake simulations with viscoelastic attenuation}, in 2016 International
  Conference on High Performance Computing \& Simulation (HPCS), pages
  908--916, IEEE, \doi{10.1109/HPCSim.2016.7568431}.

\bibitem[{Uphoff et~al.(2017)Uphoff, Rettenberger, Bader, Madden, Ulrich,
  Wollherr, and Gabriel}]{uphoff_extreme_2017}
Uphoff C, Rettenberger S, Bader M, Madden EH, Ulrich T, Wollherr S, and Gabriel
  AA (2017), Extreme scale multi-physics simulations of the tsunamigenic 2004
  {Sumatra} megathrust earthquake, in Proceedings of the {International}
  {Conference} for {High} {Performance} {Computing}, {Networking}, {Storage}
  and {Analysis}, {SC} '17, pages 21:1--21:16, ACM, New York, NY, USA,
  \doi{10.1145/3126908.3126948}, event-place: Denver, Colorado.

\bibitem[{Uski and Tuppurainen(1996)}]{Uski1996}
Uski M and Tuppurainen A (1996), {A new local magnitude scale for the {Finnish}
  seismic network}, \emph{Tectonophysics}, 261(1-3):23--37,
  \doi{10.1016/0040-1951(96)00054-6}.

\bibitem[{Verdon and Bommer(2021)}]{Verdon2021}
Verdon JP and Bommer JJ (2021), Green, yellow, red, or out of the blue? {An}
  assessment of traffic light schemes to mitigate the impact of hydraulic
  fracturing-induced seismicity, \emph{Journal of Seismology}, 25(1):301--326,
  \doi{10.1007/s10950-020-09966-9}.

\bibitem[{{Vernon} et~al.(2012){Vernon}, {Tytell}, {Hedlin}, {Walker}, {Busby},
  and {Woodward}}]{VernonEGU2012}
{Vernon} F, {Tytell} J, {Hedlin} MAH, {Walker} K, {Busby} R, and {Woodward} R
  (2012), Integration of infrasound, atmospheric pressure, and seismic
  observations with the {NSF EarthScope USArray} transportable array, in EGU
  General Assembly Conference Abstracts, EGU General Assembly Conference
  Abstracts, page 10770.

\bibitem[{Wald(2021)}]{Wald2021}
Wald DJ (2021), {Comment on “Which Earthquake Accounts Matter?” by Susan E.
  Hough and Stacey S. Martin}, \emph{Seismological Research Letters},
  93(1):500--505, \doi{10.1785/0220210072}.

\bibitem[{Waxler and Assink(2019)}]{WaxlerBook2019}
Waxler R and Assink J (2019), {Propagation Modeling Through Realistic
  Atmosphere and Benchmarking}, pages 509--549, Springer International
  Publishing, Cham, \doi{10.1007/978-3-319-75140-5_15}.

\bibitem[{Waxler et~al.(2022)Waxler, Hetzer, Assink, and Blom}]{waxler2022two}
Waxler R, Hetzer CH, Assink JD, and Blom P (2022), A two-dimensional effective
  sound speed parabolic equation model for infrasound propagation with ground
  topography, \emph{The Journal of the Acoustical Society of America},
  152(6):3659--3669.

\bibitem[{Westaway and
  Younger(2014)}]{westawayQuantificationPotentialMacroseismic2014}
Westaway R and Younger PL (2014), Quantification of potential macroseismic
  effects of the induced seismicity that might result from hydraulic fracturing
  for shale gas exploitation in the {{UK}}, \emph{Quarterly Journal of
  Engineering Geology and Hydrogeology}, 47(4):333--350,
  \doi{10.1144/qjegh2014-011}.

\bibitem[{Zhan(2019)}]{Zhan2019}
Zhan Z (2019), {Distributed Acoustic Sensing Turns Fiber‐Optic Cables into
  Sensitive Seismic Antennas}, \emph{Seismological Research Letters},
  91(1):1--15, \doi{10.1785/0220190112}.

\end{thebibliography}
\end{footnotesize}

\end{document}